\documentclass[structabstract]{aa}  

\usepackage{graphicx}
\usepackage{txfonts}

\RequirePackage{array,hyperref} 
\RequirePackage{subfig,color,pifont}
\RequirePackage{blkarray,hhline,multirow}
\RequirePackage{bold-extra}
\RequirePackage{bbm}
\RequirePackage{etoolbox}
\RequirePackage{rotating}

\definecolor{dkmauve}{rgb}{0.70,0.01,0.50}
\definecolor{dkblue}{rgb}{0.05,0.05,0.70}

\newcommand*\lephare{L\textsc{e} P\textsc{hare}~}

\hypersetup{ bookmarks=true, unicode=true, pdftoolbar=true, pdfmenubar=true, 
pdffitwindow=true, pdfstartview={FitH}, pdfauthor={SP}, 
colorlinks=true, linkcolor=dkblue,citecolor=dkblue, urlcolor=dkmauve, breaklinks }

\begin{document} 

\title{The CFHQSIR survey: a Y-band extension of the CFHTLS-Wide survey \thanks{The CFHQSIR catalogue for 8.6 million sources down to $[(\mathrm{z'} \leq 23.5) \lor (\mathrm{Y} \leq 23.0)]$ is available at the CDS via anonymous ftp to  \url{cdsarc.u-strasbg.fr} (130.79.128.5)
		or via \url{http://cdsweb.u-strasbg.fr/cgi-bin/qcat?J/A+A/}}\fnmsep\thanks{The CFHQSIR catalogue is also available at: \url{http://apps.canfar.net/storage/list/cjw/cfhqsir}, as well as the Y-band images.}}

\author{S. Pipien \inst{1}
	\and S. Basa \inst{1}
	\and J.-G. Cuby \inst{1}
	\and J.-C. Cuillandre \inst{2}
	\and C. Willott \inst{3}
	\and T. Moutard \inst{4}
	\and J. Chatron \inst{1}
	\and S. Arnouts \inst{1}
	\and P. Hudelot \inst{5}
}
\institute{Aix Marseille Univ, CNRS, LAM, Laboratoire d'Astrophysique de Marseille, Marseille, France\\
	\email{sarah.pipien@lam.fr}
	\and
	CEA/IRFU/SAp, Laboratoire AIM Paris-Saclay, CNRS/INSU, Université Paris Diderot, Observatoire de Paris, PSL Research University, F-91191 Gif-sur-Yvette Cedex, France
	\and
	NRC Herzberg, 5071 West Saanich Rd, Victoria, BC V9E 2E7, Canada
	\and
	Department of Astronomy \& Physics and Institute for Computational Astrophysics, Saint Mary's University, 923 Robie Street, Halifax, Nova Scotia, B3H 3C3, Canada
	\and
	Institut d’Astrophysique de Paris, 98bis Boulevard Arago, F-75014 PARIS, France
}
 
\abstract
{The Canada-France-Hawaii Telescope Legacy Survey (CFHTLS) has been conducted over a five-year period at the CFHT with the MegaCam instrument, totaling 450 nights of observations. The Wide Synoptic Survey is one component of the CFHTLS, covering 155 square degrees in four patches of 23 to 65 square degrees through the whole MegaCam filter set (u*, g', r', i', z') down to {i'$_\mathrm{AB}$} = 24.5. }
{With the motivation of searching for high-redshift quasars at redshifts above 6.5, we extend the multi-wavelength CFHTLS-Wide data in the Y-band down to magnitudes of $\sim$ 22.5 for point sources (5$\sigma$) .}
{We observed the four CFHTLS-Wide fields (except one quarter of the W3 field) in the Y-band with the WIRCam instrument (Wide-field InfraRed Camera) at the CFHT. Each field was visited twice, at least three weeks apart. Each visit consisted of two dithered exposures. The images are reduced with the Elixir software used for the CFHTLS and modified to account for the properties of near-InfraRed (IR) data. Two series of image stacks are subsequently produced: four-image stacks for each WIRCam pointing, and one-square-degree tiles matched to the format of the CFHTLS data release. Photometric calibration is performed on stars by fitting stellar spectra to their CFHTLS photometric data and extrapolating their Y-band magnitudes.}
{After corrections accounting for correlated noise, we measure a limiting magnitude of Y$_\mathrm{AB} \simeq 22.4$ for point sources (5$\sigma$) in an aperture diameter of 0\farcs93, over 130 square degrees. We produce a multi-wavelength catalogue combining the CFHTLS-Wide optical data with our CFHQSIR (Canada-France High-z quasar survey in the near-InfraRed) Y-band data. We derive the Y-band number counts and compare them to the Vista Deep Extragalactic Observations survey (VIDEO). We find that the addition of the CFHQSIR Y-band data to the CFHTLS optical data increases the accuracy of photometric redshifts and reduces the outlier rate from 13.8\% to 8.8\% in the redshift range 1.05 $ \lesssim$ z $ \lesssim$ 1.2. }
{}


\maketitle
\titlerunning{CFHQSIR}

\section{Introduction}
The Canada-France-Hawaii Telescope Legacy Survey (CFHTLS) has been conducted from mid-2003 to early 2009 at the CFHT (Canada-France-Hawaii Telescope) using the MegaCam wide field imaging camera, and totaling 450 nights (2300 hours) of observations. MegaCam is a $1\degr \times 1\degr$ field of view 340 megapixels camera \citep{boulade} installed at the prime focus of the 3.6m CFH telescope.

The CFHTLS consists of three distinct survey components: the supernovae and deep survey (SNLS and the Deep Survey), a wide synoptic survey (the Wide Survey) and a very wide shallow survey (the Very Wide Survey). The Wide Survey covers 155 square degrees in four patches of 23 to 65 square degrees through the whole filter set (u*, g', r', i', z') down to an AB magnitude i'=24.5.

As part of a large programme aimed at searching for quasars at redshifts of $\sim 7$, we carried out Y-band near-IR observations of the CFHTLS-Wide Survey. This survey is intended to extend to higher redshifts the highly successful $5.8 < z < 6.5$ Canada-France Quasar Survey \citep{willott1,willott2,willott3}, from which our survey takes its name - CFHQSIR. 

This data paper describes the CFHQSIR data. In the next section, we describe the survey observations. In Sect. \ref{sec:processing}, we present the data pre-processing, the photometric calibration and the CFHQSIR data format. In Sect. \ref{sec:properties} we discuss the main properties of the CFHQSIR data in terms of image quality and limiting magnitude. Section \ref{sec:catalogue} addresses preliminary analyses (number counts and photometric redshifts) regarding the CFHQSIR data and the added value of the Y-band magnitude.

\section{The CFHQSIR observations}

The Wide-field InfraRed Camera (WIRCam) is the near-IR mosaic imager at the CFHT, which has been in operation since November 2006. WIRCam complements the one-square-degree optical imager, MegaCam, which has been in operation at CFHT since 2003. 

WIRCam is mounted at the prime focus of the  3.6m CFH telescope. It is equipped with an image stabilization unit, which consists of a tip-tilt glass plate in front of the camera activated by the signal read-out from small 14$\times$14 pixel regions centred on bright stars \citep{puget}. The camera cryostat includes the eight-lens field corrector and an entrance window. The detector focal plane consists of a mosaic of four 2k $\times$ 2k HAWAII2-RG detectors separated by 7 mm. The sampling on sky is 0\farcs3 per 18 $\mathrm{\mu}$m pixel, providing a total field of view of 20\arcmin $\times$ 20\arcmin\ with a 2 \arcmin\-wide central cross between the four detectors.

\begin{figure}\centering
      \includegraphics[width=0.48\textwidth]{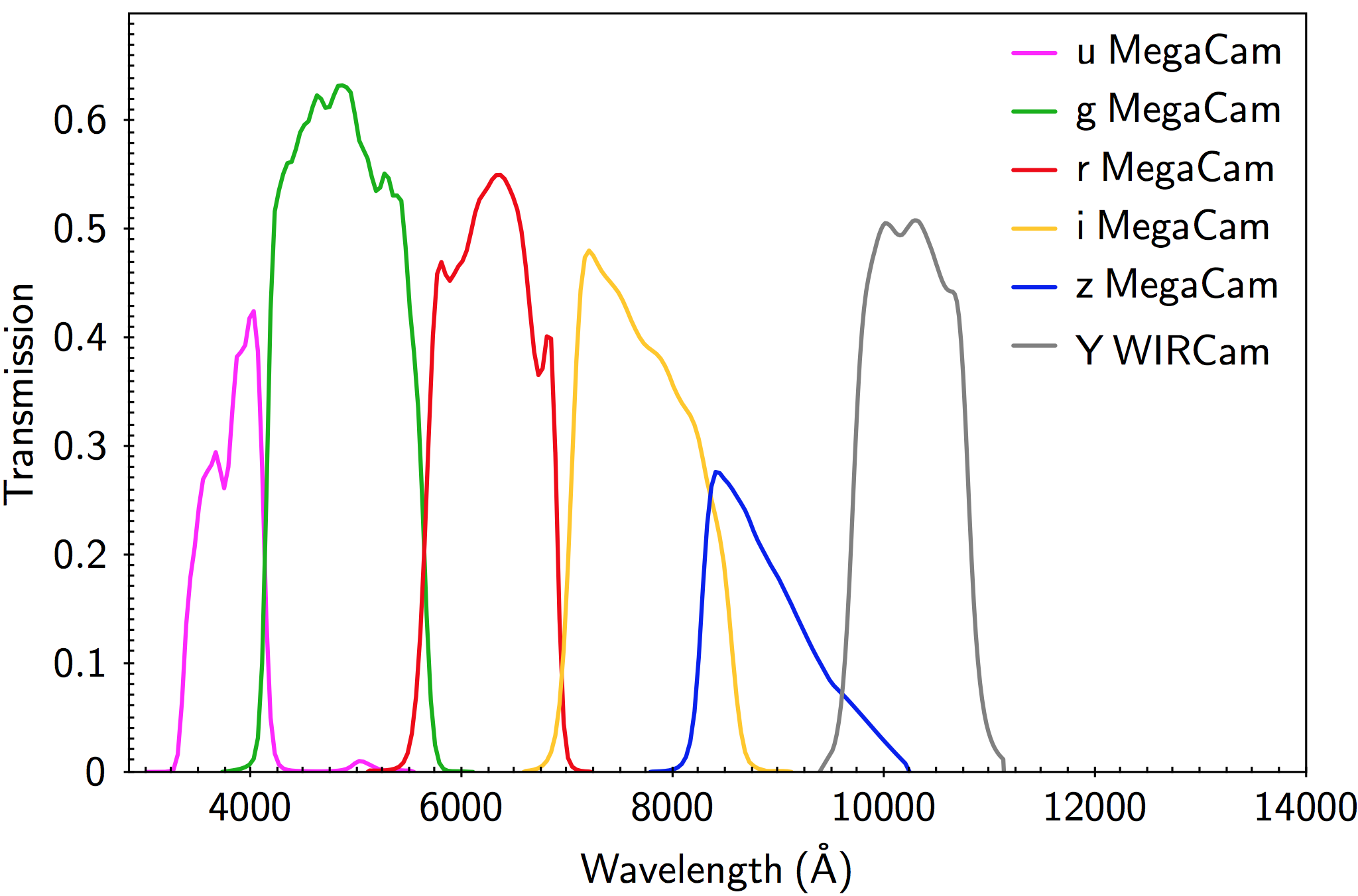}
    \caption{Instrument and telescope total efficiency (optics and detector) in the MegaCam u*, g', r', i', z' filters and WIRCam Y-band.}
  \label{fig:filtersWircam}
\end{figure}

The transmission curve of the Y-band filter used in this work is shown on Fig.~\ref{fig:filtersWircam}, together with the MegaCam filter transmission curves.

\begin{figure*}[ht!]
	\centering
   	    	 \includegraphics[width=0.75\textwidth]{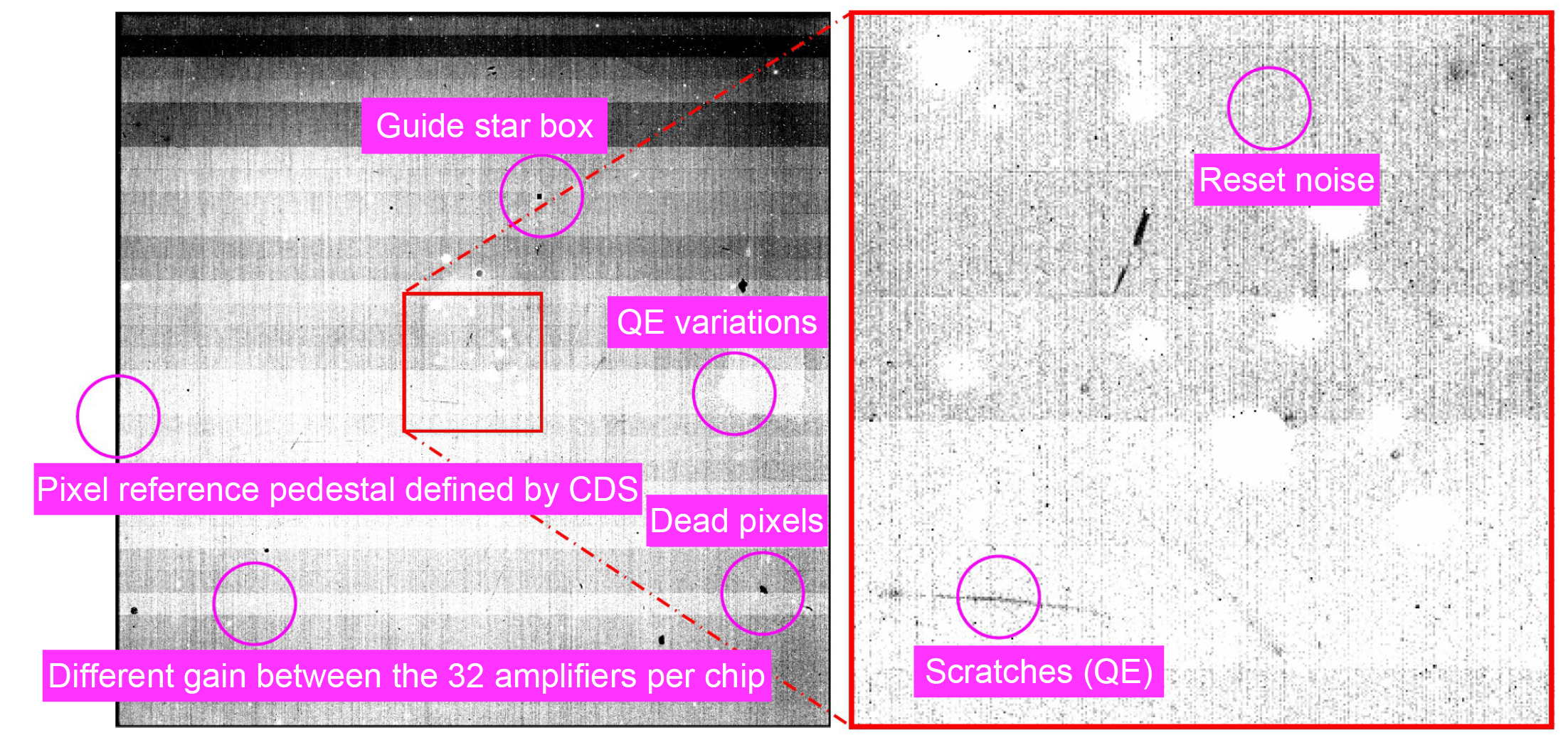}
  	\caption{WIRCam detector features.}
  \label{fig:RawDataLook}
\end{figure*}

The survey observations consisted of four dithered 75-seconds Y-band images of the same field. They were split into two visits, with dithering within a visit and between visits of 2\arcsec. The two visits were separated by at least 20 days in order to discard slow moving objects and strongly variable objects. The time difference between visits of the same field varied strongly between the four CFHTLS fields. For W1 and W4, the observations were performed very rapidly at the beginning of the survey, while for W2 and W3 (winter and spring fields), they suffered from poorer weather conditions and were executed over much longer periods of time. The interval between visits of the same fields were in the range of 20-40 days for 50\% (W1) and 90\% (W4) of the observations, whereas for the W2 and W3 fields, most of the visits were separated by approximately one year. 

The mapping of the fields was organized so as to match the CFHTLS one-square-degree tiles, with nine WIRCam pointings per CFHTLS tile. The observations were defined so as to prevent the execution of the second-visit observations in the same sequence as for the first visits, which could have generated similar persistence patterns in the data.

A total of approximately 150 hours of observations were conducted during four CFHT semesters, 2010B, 2011B, 2012A, and 2012B. The observations were carefully checked by our team in the days following the observations using the Elixir-IR pre-processing package (see next section), and observations that were executed under high background conditions and/or poor seeing conditions were not validated and were subsequently re-executed.

\section{Data processing}
\label{sec:processing}
\subsection{Need for a custom pipeline: Elixir-IR} \label{sec:custom}
The CFHT offers to all its WIRCam users a data detrending and calibration service, the `I`iwi pipeline \citep{thanjavur}, which aims to remove the instrumental signature and delivers an absolute photometric calibration within 5\%. This pipeline relies on generic recipes such as acquiring flat fields over a whole observing run in order to create master detrending frames that are applied to all frames acquired during that observing period. This pipeline is directly derived and inspired from the earlier Elixir pipeline \citep{magnier} developed at the CFHT for its workhorse wide field imaging instrument operating in the visible domain, MegaCam. However, experience with Elixir has shown that particular attention to effects such as illumination and stability of the photometric response \citep{regnault} is required when ultimate photometric stability is sought.
	
	We therefore initiated a study of the stability in time of the WIRCam data. With an objective of a few percent photometric accuracy across the field of view, an average flat field gathered over a seven to 15 day period was found clearly inadequate since the instrument exhibits changes in its response from night to night and especially from twilight to night. Indeed, dividing flat fields acquired over several days clearly shows residuals at small and large scales at the 5\% level. These residuals affect the photometry at both medium and large scales.

	Since ground-based near-IR observations are quickly sky- background-dominated, even in the Y-band CFHQSIR short exposures, it is possible to build flat-field frames directly from the science data over reasonably short timescales, thereby limiting the effects of instrumental variations over days. We therefore developed a dedicated version of the Elixir pipeline, called Elixir-IR, aimed at minimizing the specifics of the WIRCam detectors, and more generally of near-IR detectors. The various processing steps are described in the following sub-sections.

\subsection{Detector and raw data properties}
A single 2048 $\times$ 2048 pixel detector array is read out through 32 outputs \citep{puget} in less than four seconds. The array is surrounded on the left, right, top and bottom by four columns and four lines of reference pixels, which do not integrate light and that are used to calibrate the additive level in the image signal (pedestal) as well as gradients and reset level from column to column, as described below.

A close inspection of a raw image uniformly illuminated shows important features that need to be accounted for, either by masking pixels that are not suitable for science, or by proper detrending with additive, non-linearity, and multiplicative corrections (see Fig.~\ref{fig:RawDataLook}).  We list here the major features: a) each readout amplifier has its own electronic operating properties leading to different gains (32 horizontal bands), b) the pixel reference pedestal can be sampled on the pixels on the left and right of the array, c) WIRCam using on-chip guiding, the pixels in the guide box (20 $\times$ 20 pixels) must be rejected from the scientific analysis\footnote{The columns and rows aligned to the guiding window are, however, not rejected.}, d) the quantum efficiency shows variations, even scratches, across the surface, a result of the manufacturing process, e) there are non-responsive pixels across all detectors, either isolated or in clusters of varying sizes, f) the electronic reset noise when reading out the detector through the multiplexer causes a non-constant pedestal setting, seen as vertical comb structures across the entire height of the image. 

In the following, we describe how each of those features is handled in the Elixir-IR pipeline per detector and in a time-sequential approach. The pipeline handles all four detectors in parallel in a similar fashion.

\begin{itemize}
\item {\bf Reset noise}.
  A single reset signal is applied to the entire detector once all columns have been readout through the 32 amplifiers (each readout stripe being 64 pixels high). There is a noise associated to this process and this sets a new pedestal for the next column readout, an effect than can be as high as 10 to 15 ADUs (Analog-to-Digital Units). This effect is clearly visible on low background frames. In practice, this reset noise is highly correlated over three to four columns. The four lines of reference pixels at the top and bottom of the array are equally affected and can be used to build a one-dimensional horizontal model that is subtracted to the entire image. Figure~\ref{fig:Reset} shows an image with and without the reference pixel correction.

\begin{figure}\centering
	\includegraphics[width=0.3\textwidth]{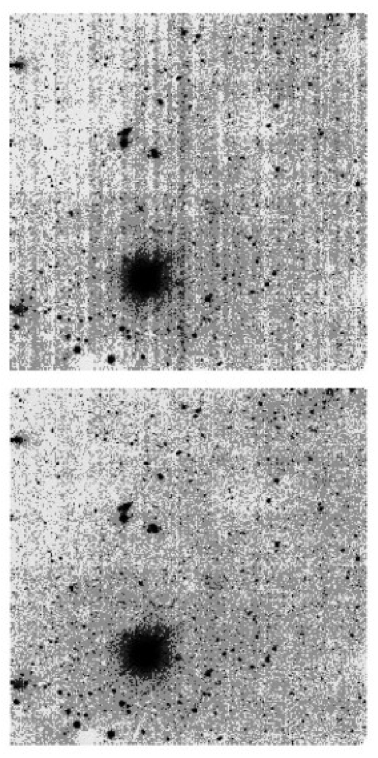}
	\caption{Illustration of the reference pixel correction (top: without, bottom: with) for an image of the W4 field.} 
	\label{fig:Reset}
\end{figure}

\item {\bf Frame pedestal}.
  The four columns on the left and right of the detector, now corrected from the reset noise, show a constant level, an artificial pedestal set by the readout electronics to ensure that no signal ever gets to negative levels after double correlated sampling. On WIRCam this artificial level is set around 1,000 ADUs. A simple median of all pixels from those eight columns is subtracted.

\item {\bf Dark current}.
  The WIRCam cryostat keeps the detectors at very low temperature (80K) and the dark current of 0.8 electrons per second and per pixel is nearly negligible with respect to the typical sky background level in the Y-band. Although stable over time, the dark current is, however, strongly dependent on integration time. Dark frames at the exposure time of the CFHQSIR (75 seconds) were taken in dark conditions to produce high signal-to-noise dark frames to substract from the science data.

\item {\bf Non-linearity}.
  The gain, readout noise, non-linearity, and saturation limits of all four detectors were derived using the photon transfer function \citep{Janesick}. The detector gains are of the order of 4.0 electron per ADU (all four are similar within 0.2 electrons per ADU), the readout noise is equal to 28 electrons, and the saturation limit varies from 100,000 electrons to 160,000 electrons between detectors.

\item  {\bf Bad pixels}.
  We took the approach of masking all pixels that have a strongly discrepant response relative to the average response of the arrays. We compared two high signal-to-noise flat-field frames, one with an integration time twice as long as the other. After scaling with the integration time and subtraction of the two images, we rejected all pixels with a residual deviating from zero by more than 1\% of the signal. The final bad pixel count is of the order of 2\% per detector. The windows used by WIRCam for on-chip guiding (400 pixels) are also treated as bad pixels on an image by image basis by Elixir-IR.

\item {\bf Building a flat field}.
Near IR observations are characterized by strong spatial and temporal sky variations (OH airglow), as well as detector property variations (e.g. gain) over time, possibly due to controller temperature variations. Building appropriate flat fields therefore requires caution.

We initially performed a number of tests to check the stability of the illumination pattern over time. We found that there are gain drifts of $\approx$ 1\% over timescales of the order of three hours. This makes the use of twilight flat fields inappropriate, not considering the strong spatial variations of the twilight flat-field patterns which can exceed 5\% over short timescales. In order to limit the gain variations to well below 1\%, we adopted time windows of the order of 30 minutes to generate sky flat fields.

Considering the CFHQSIR observing strategy, and including readout and telescope pointing overheads, 14 exposures (seven dithered pairs of 75s each) fit within an $\approx$ 20 minute long time window. The signal is largely dominated by the high sky background (typically 2000 to 5000 ADUs) ; 14 exposures on seven independent pointings are therefore adequate to derive good flat fields. The flat-field frames are generated by averaging the 14 frames with iterative sigma clipping making use of the detector gain and noise properties. We note that including the central science frame in the flat-fielding procedure can introduce a photometric bias, depending on the averaging method. This bias can be as high as the inverse of the number of frames when using a simple median. However, with our iterative image rejection method based on photon noise properties, the bias shrinks to negligible levels (less than 1 \%).

\end{itemize}

\subsection{Inspection and stacking}

After removal of the detector features and flat-fielding, the images are free of cosmetics patterns. The main remaining patterns are cosmic rays and low frequency background residuals (tilts) that are due to the structure of the OH airglow and to its evolution over time. Except for these residuals, all features on the science images are flattened at the 0.1\% level (min-max), comparable to what is achieved on CCDs using the Elixir-LSB pipeline on MegaCam data.

Once the individual frames are fully detrended, they are visually inspected through a quick preview procedure to check the images look fine and in particular if the sky background (airglow) has a normal behavior, changing in subtle ways from one exposure to the next (see bottom image of Fig.~\ref{fig:Reset} for an illustration of the typical appearance of an Elixir-IR detrended frame, all four detectors being normalized to the same response). Individual frames are then calibrated for astrometry using the Two Micron All Sky Survey (2MASS) catalogue \citep{Skrutskie}. 

\begin{figure}\centering
	\includegraphics[width=0.5\textwidth]{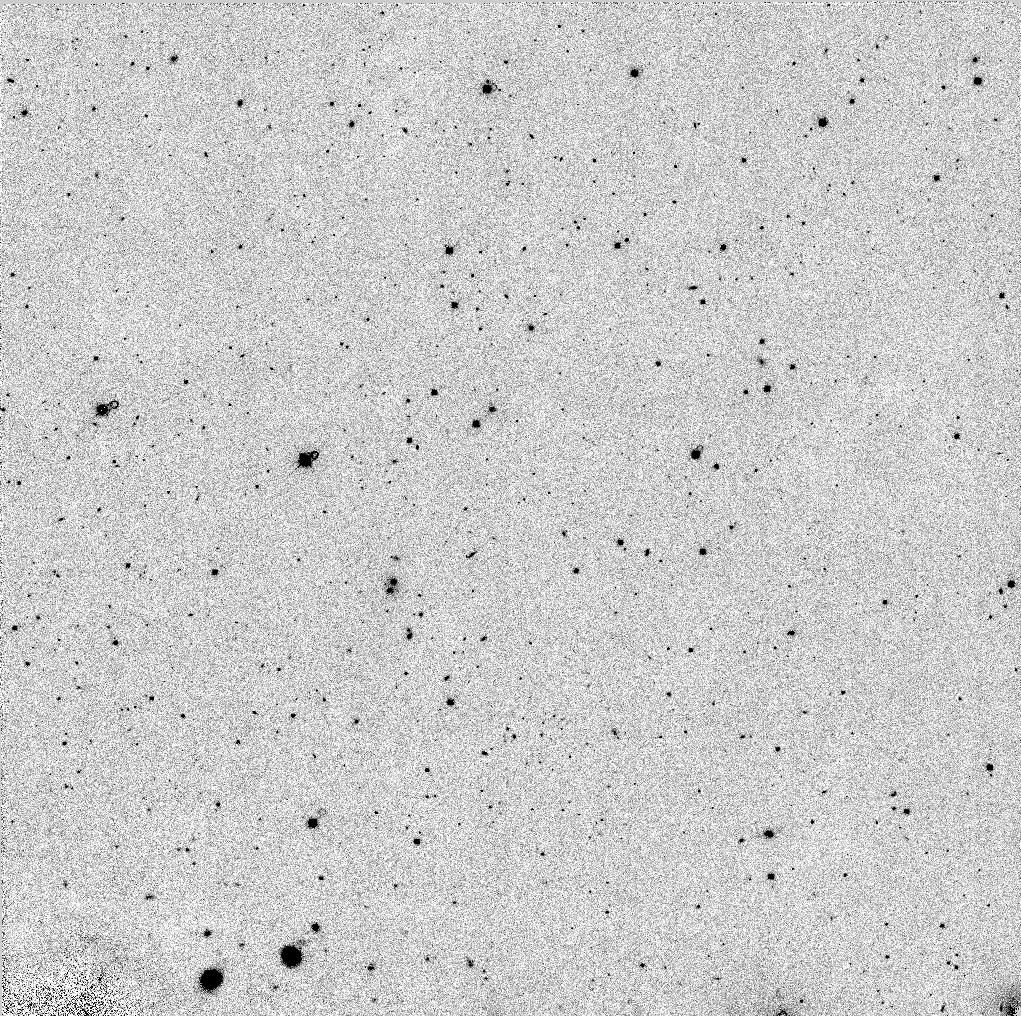}
	\caption{Image of one quadrant of a WIRCam stack in the W4 field produced by the Elixir-IR pipeline.}
	\label{fig:Elixir-1}
\end{figure}

For stacking, we adopt the AstrOmatic suite\footnote{\url{www.astromatic.net}} by E. Bertin to derive a fine astrometry (SCAMP) and resample the images for alignment on a given grid and stacking (SWarp). The sky background is subtracted as two-dimensional (2D) plane fits to individual images. In most cases, stacks consist of four images. Despite this small number of data points, good rejection of bad pixels, cosmic rays, and spurious signals is achieved with sigma-clipping using the detector noise parameters (gain and noise). Remnant signals from previous observations of heavily saturated stars are mostly removed during this stacking process but can be easily identified on averaged stacks. If the remnant originates from an image recorded earlier in the night and unrelated to CFHQSIR, the persistence pattern consists of two spots along the dither direction, whereas the pattern has three spots if the remnant originates from a dithered pair of previous CFHQSIR observations. Due to the small offsets between images, it was not necessary to apply illumination correction (see next section) before stacking. Weight maps for each stack are also produced by SWarp during the resampling and stacking process. An example is shown in Fig.~\ref{fig:Elixir-1}. These weight maps include the 2 \arcmin\ gaps between the four detector arrays. The four-image stacks so produced will be referred to in the following as WIRCam stacks.

\subsection{Photometric calibration}
\label{sec:calphot}

Information on photometric calibration in the Y-band is scarce due to the limited use of this band. Direct photometric calibration of the Y-band using Vega-like A0 stars has been performed by for example \citet{Hillenbrand}. The Y-band UKIDSS (UKIRT Infrared Deep Sky Survey) - WFCAM (Wide Field Camera) data were anchored to the Two-Micron All-Sky Survey (2MASS) and calibrated in the Vega system by zeroing the colours of blue stars present in the data \citep{hodgkin}.

Considering the connection between the CFHQSIR and CFHTLS datasets, we chose to anchor the CFHQSIR photometric calibration to the CFHTLS and to treat the Y-band extension as an extrapolation in wavelength. For convenience, we decided to perform the photometric calibration directly on the image stacks (WIRCam stacks) as described in the previous section. To this end, we select unsaturated and high signal-to-noise ratio (SNR) stars in our Y-band catalogue, and we fit stellar spectra to their griz magnitudes, making additional use, when available, of 2MASS JHK photometric data.

In more detail, we proceed as follows:

\begin{enumerate}
\item From our Y-band catalogue, we select unsaturated objects classified as stars by SExtractor and with an SNR matching our target photometric accuracy of a few percent. In practice, we selected objects with SNR $>$ 40.
\item We search for these stars in the CFHTLS catalogue within a 1\arcsec\ search radius and we use the ``IQ20" magnitudes, following the prescription of  \citet{cfhtls} for point-like objects. The CFHTLS ``IQ20'' magnitudes are the true total magnitudes integrated over an aperture 20 times the full width at half maximum (FWHM) of the point spread function (PSF).
\item When available, the 2MASS (profile-fitting) magnitudes in the J, H, and K$_s$ bands are also used in the fits. 
\item For each star, we fit the optical photometric data with the \citet{Pickles} library of stellar spectra and we derive the Y-band magnitude from these fits. To this aim, we use the \lephare \citep{Arnouts1999, Ilbert2006} photometric redshift sofware.
\item We intentionally use a limited number of spectra representative of the most common stars likely to be present in our samples, and we exclude cold stars that have broad absorption features in the near-IR, therefore preventing a reliable extrapolation of their spectra. We therefore limit the spectral types used for our fitting procedure from G0 to K7. We also limit the photometric bands used in the fit to the riz[JHKs] bands to avoid the sensitivity to metallicity of the u and g bands. We exclude poor fits as measured by the \lephare internally derived $\chi^2$ values.
\end{enumerate}

\begin{figure}\centering
    \includegraphics[width=0.5\textwidth]{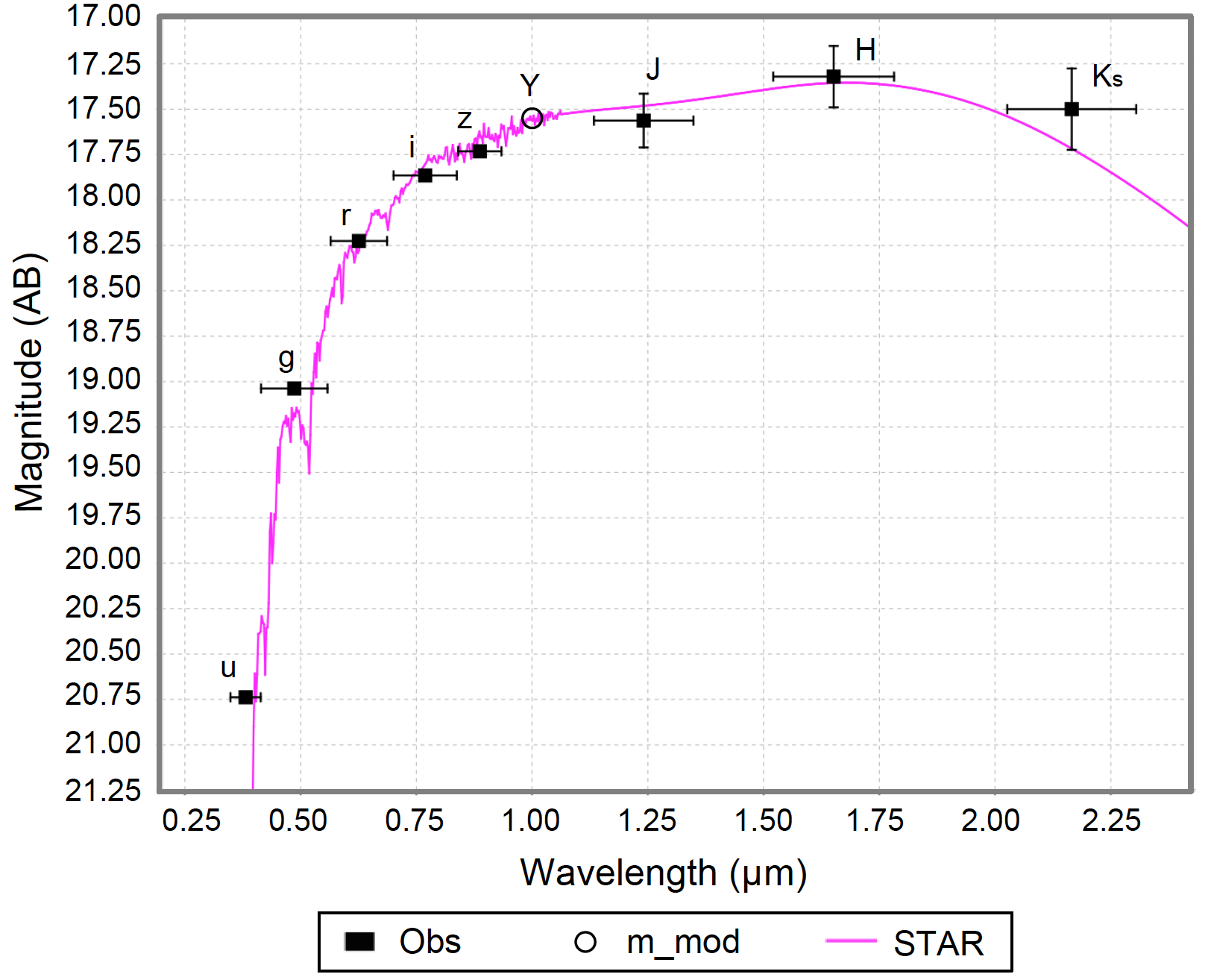}
  \caption{Example of a fit with a  K-type spectrum in a case where 2MASS data are available. The derived Y-band magnitude is indicated by the circle.}
  \label{fig:sed_fit}
\end{figure}

An example of a fit with a stellar spectrum to the spectral energy distribution of a star  for which there are ugriz CFHTLS and JHKs 2MASS photometry is shown in Fig.~\ref{fig:sed_fit}. The u and g bands data points are shown but are not used in the fit. In total, we use 45,500 stars to perform this photometric calibration over the four CFHQSIR fields, leading to an average photometric zero point per quadrant (detector) of each WIRCam stack. We tentatively observe a less than 1\% difference between the zero points derived with or without 2MASS data, not significant enough to be corrected for.

\begin{figure}\centering
	\includegraphics[width=0.5\textwidth]{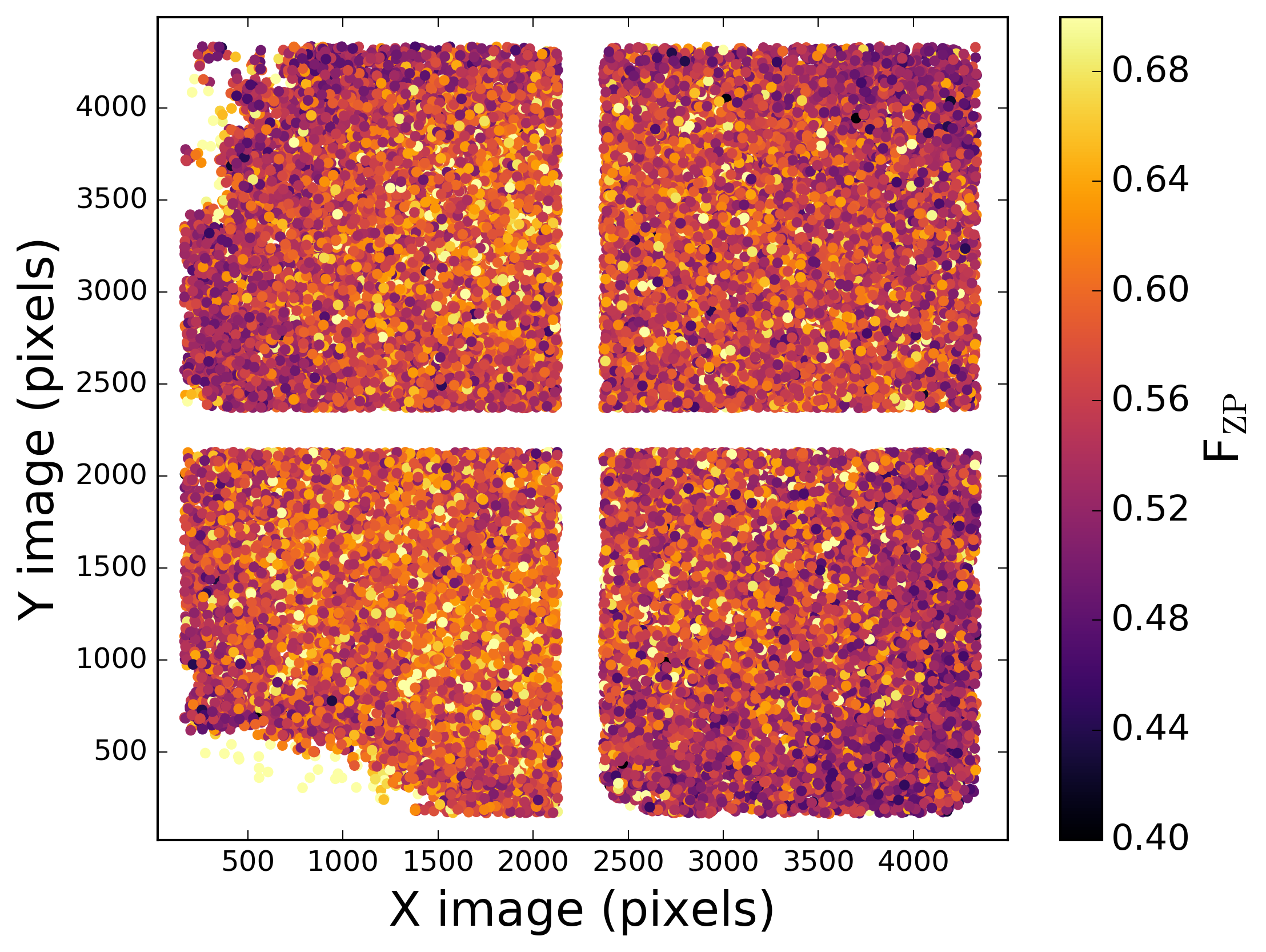}
	\caption{Colour-coded flux calibration coefficients per star $\mathrm{F_{ZP}}$ as a function of the star position in the WIRCam focal plane.The calibration coefficient $\mathrm{F_{ZP}}$ is defined as the ratio, before calibration, of the flux fitted with Le Phare to the flux measured by SExtractor.}
	\label{fig:MapCoeffZPFlux}
\end{figure}

\begin{figure}\centering
	\includegraphics[width=0.5\textwidth]{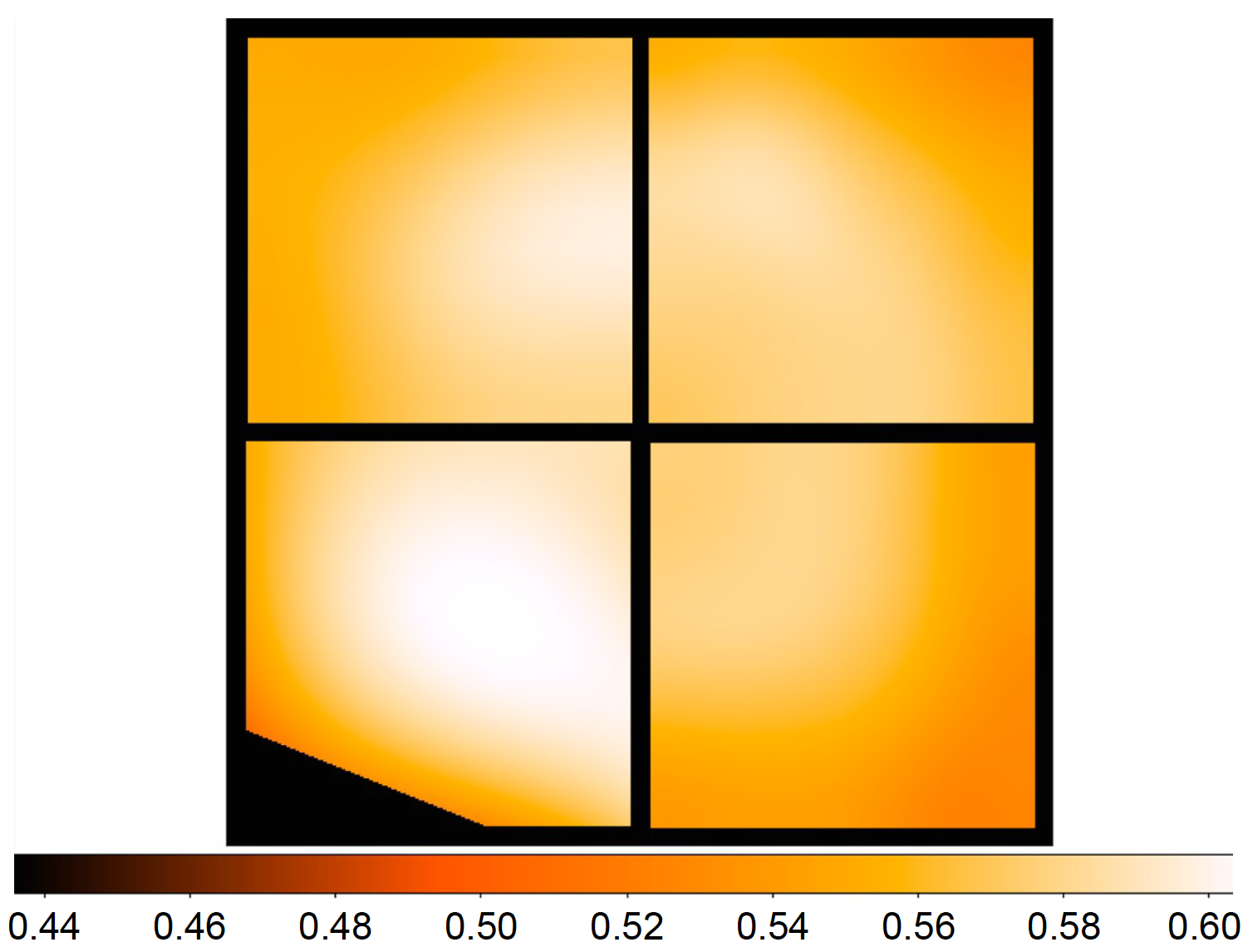}
	\caption{Map used to correct the spatial non-uniformity of the photometric response. The peak-to-peak relative amplitude variation in this image is 20 \%.}
	\label{fig:illumination}
\end{figure}

After this first pass the relative dispersion of the calibration coefficients is equal to 10\%. This dispersion includes photometric variations between stacks and spatial variations within one quadrant and between quadrants. Spatial variations of the zero points due to optical distortions and/or sky concentration are introduced during the flat-fielding process \citep[see e.g.][]{regnault}. These variations can be seen on Fig.~\ref{fig:MapCoeffZPFlux}, which shows the zero points projected onto a single WIRCam image in pixel coordinates. From this image, we generate a 2D fit per quadrant (Fig.\ref{fig:illumination}), usually referred to as illumination-correction map. This map is subsequently used to correct all the WIRCam stacks. 
 
After correction for the illumination map, we re-compute all the zero points for the whole dataset. For each WIRCam stack, we derived the final zero point by averaging with sigma-clipping rejection the zero points measured for each star present in the stack. We require that a minimum of six zero points be measured in each stack. When there are less than six zero points, which happened for nine stacks only over a total of 1445), we assigned the average zeropoint of the corresponding CFHTLS field. For these nine stacks, this may represent an additional photometric error of less than 10\%.

The illumination correction reduced the overall dispersion of the zero points from 10\% to 7\%, consistent with the amplitude of the illumination correction map. Finally, for consistency with the CFHTLS data \citep{cfhtls}, we normalize each stack by setting the zero point to a value of 30.0 in the AB system.

\subsection{Registration to the CFHTLS format}
\label{sec:registrationCFHTLS}
For ease of use in connection with the CFHTLS data, we decided to produce one-square-degree image tiles similar to the CFHTLS data format, in addition to the WIRCam stacks. We use SCAMP \citep{scamp} to determine the geometrical transformation between the reference z-band CFHTLS tile and the nine WIRCam stacks corresponding to this tile. In practice, the transformation is determined for each of the four quadrants of the WIRCam stacks using a three-degree polynomial. We then use SWarp \citep{swarp} to apply these transformations using a LANCZOS3 (Lanczos-3 6-tap filter)  interpolation kernel and to reformat the whole dataset to the CFHTLS tile format.

When there is overlap between adjacent WIRCam stacks within a tile, all the pixels in the overlapping regions are used. Conversely, pixels in overlaps between tiles\footnote{Adjacent CFHTLS tiles overlap by a few arcminutes.} are treated independently.

\section{Main properties of the survey}
\label{sec:properties}
We present in this section the main properties of the CFHQSIR survey data. We perform the analysis either on the WIRCam stacks or on the one-square-degree images mapped into the CFHTLS format (tiles) described in the previous section. We further divide each tile into 3 $\times$ 3 sub-images corresponding to the footprint of each WIRCam stack.

\subsection{Image quality}
\begin{figure}\centering
	\includegraphics[width=0.5\textwidth]{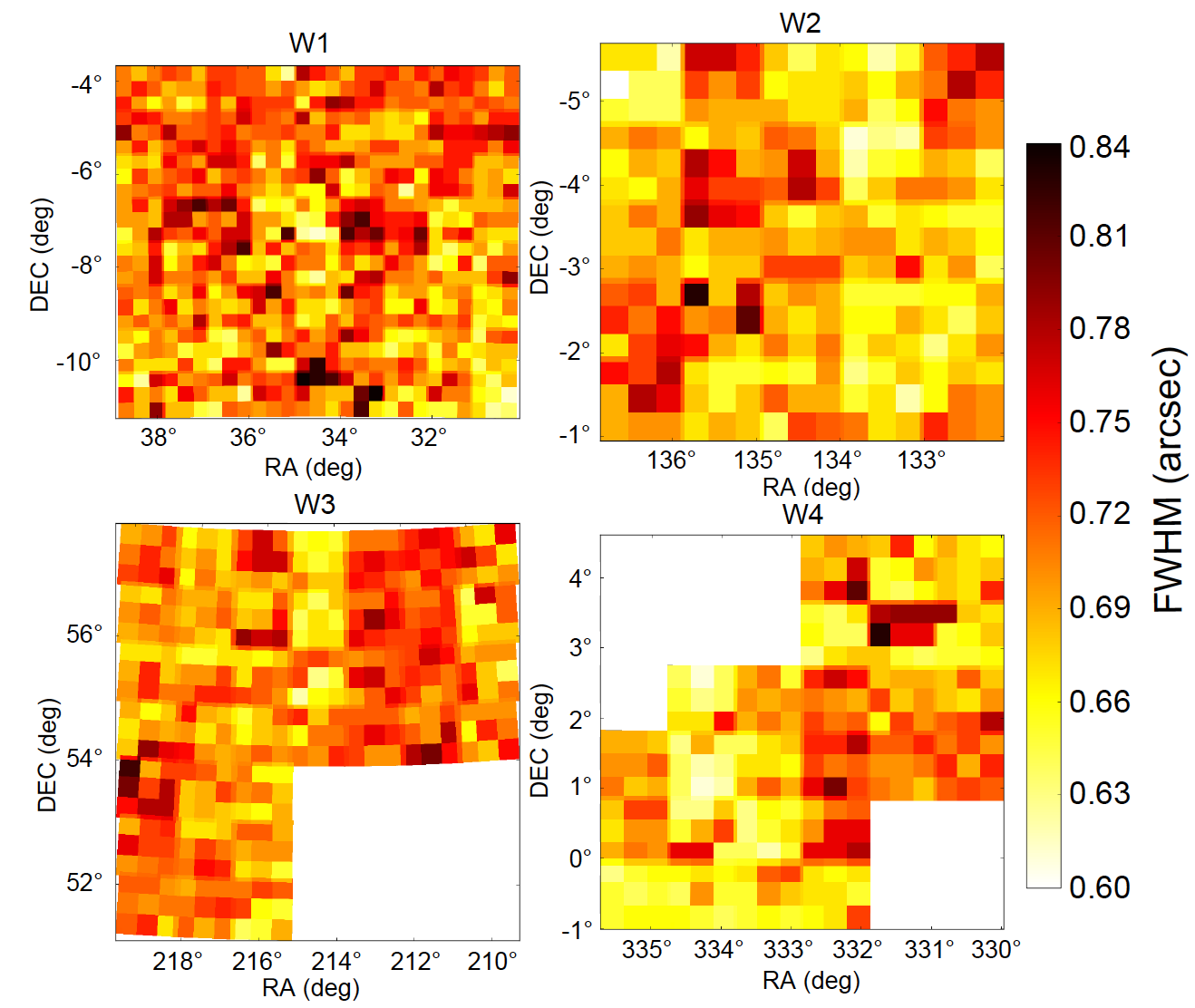}
	\caption{Maps of image quality over the four CFHTLS fields. One square or rectangle of uniform colour corresponds to the footprint of each WIRCam stack.}
	\label{fig:iq_maps}
\end{figure}

\begin{figure}\centering
    \includegraphics[width=0.5\textwidth]{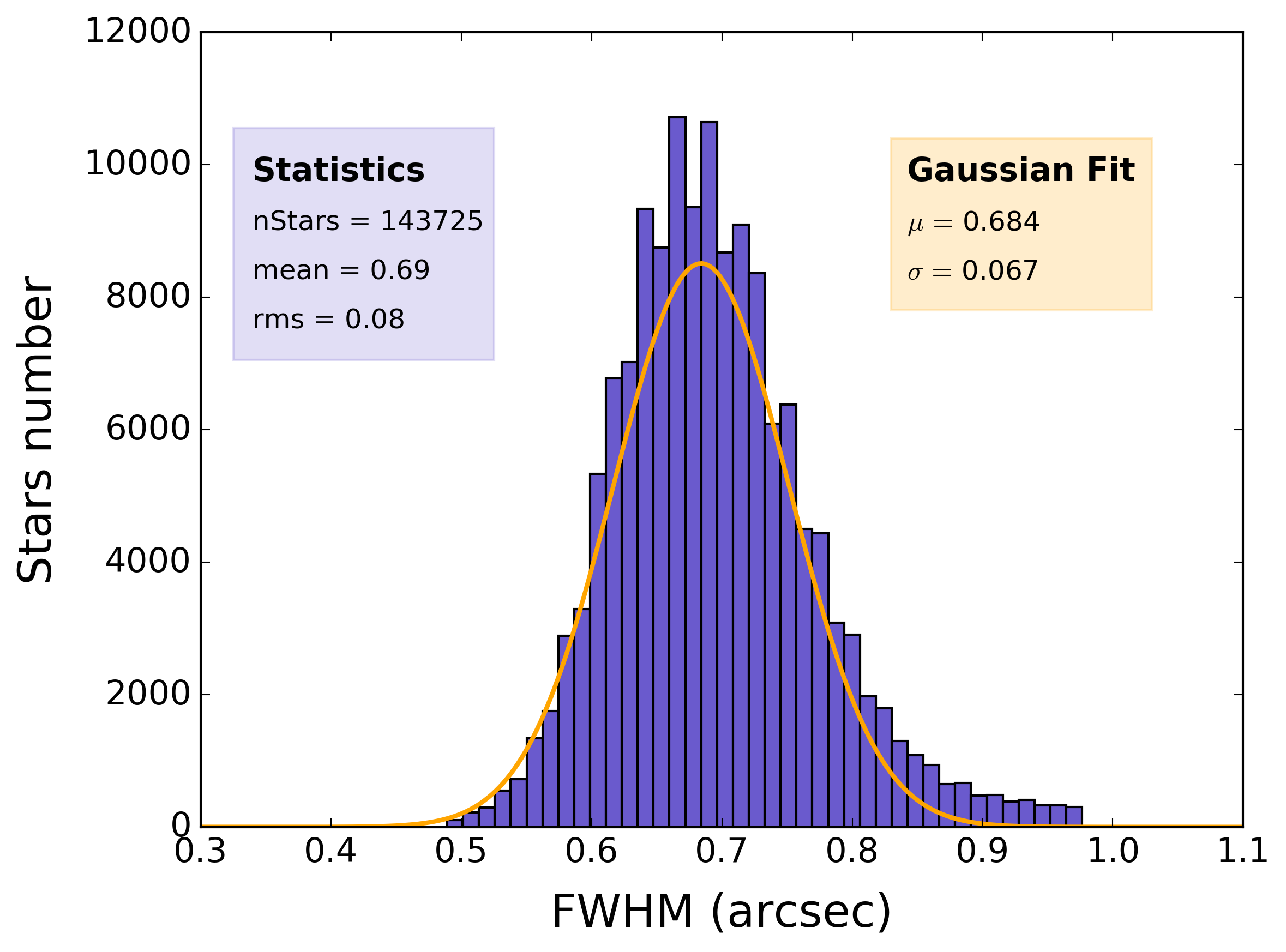}
  \caption{Histogram of the image quality over the whole survey data.}
  \label{fig:figures/iq_histo.png}
\end{figure}

All observations were carried out in service mode, with an IQ constraint in the K-band of $0\farcs55$ to $0\farcs65$, which translates, assuming a seeing limited image, to $0\farcs65$ to $0\farcs75$ in the Y-band. This is very consistent with our measured image quality. Figure~\ref{fig:iq_maps} shows maps of the median image quality per sub-tile image measured on unsaturated stars over the four CFHTLS fields. The number of stars used to determine the image quality was of the order of one hundred per sub-tile image. The histogram of the FWHM of all the stars used is represented in Fig.~\ref{fig:figures/iq_histo.png}.

\begin{figure}\centering
	\includegraphics[width=0.5\textwidth]{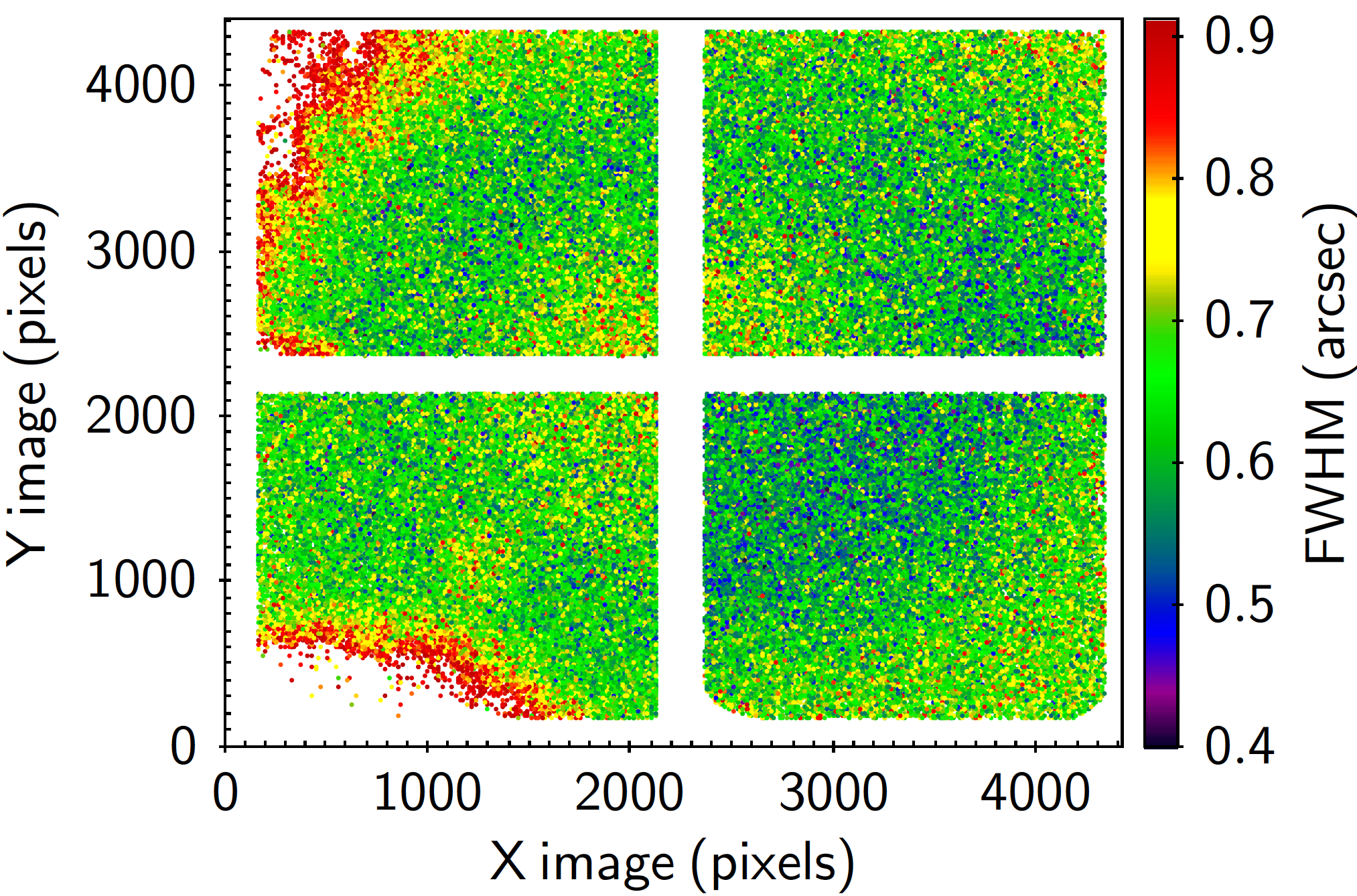}
	\caption{Image quality as a function of the X and Y pixel coordinates. Only the region at the bottom left corner of the image was masked, where poor image quality coincides with very poor detector cosmetics. The second region at the top left corner of the field of view showing poor image quality but good detector cosmetics was kept.}
	\label{fig:iq_1image}
\end{figure}

We have explored the variation of the image quality over the WIRCam images. Figure~\ref{fig:iq_1image} shows the x and y position of all unsaturated stars used for the photometric calibration (see Sect. \ref{sec:calphot}). Despite the variations in seeing, one can note the spatial variations across the WIRCam field of view. The image quality is generally worse at the edge of the field. A region at the bottom left corner of the image shows very poor image quality, attributed to detector issues since this region is adjacent to a region of very poor cosmetics and dead pixels. This bad quality region has been subsequently masked in the CFHQSIR data.

\subsection{Sensitivity}

\subsubsection{Correlated noise correction} \label{sec:error}
Image resampling introduces correlations in the noise that should be accounted for in the photometric error budgets. Discussions of this effect can be found for instance in \citet{Casertano}, \citet{Labbe} (Sect. 4.4 and equation 3 therein), \citet{Grazian} and \citet{Clement}.

Noting $\Phi_\mathrm{SE}$, the signal measured by SExtractor on an aperture of $N_\mathrm{pix}$ pixels, the photometric error measured by SExtractor in the background dominated regime can be written as

\begin{equation}
  \sigma_{\Phi_\mathrm{SE}} = \sigma_\mathrm{pix} \times \sqrt{N_\mathrm{pix}}
  \mathrm{,}
  \label{equ:error1}
\end{equation}

where $\sigma_\mathrm{pix}$ is the local pixel to pixel standard deviation of the background signal.

To take into account the correlated noise, the photometric error will be multiplied by a corrective excess noise factor $f_\mathrm{corr}$:
\begin{equation}
  \sigma_{\Phi} = f_\mathrm{corr} \times \sigma_{\Phi_\mathrm{SE}} = f_\mathrm{corr} \times \sigma_\mathrm{pix} \times \sqrt{N_\mathrm{pix}}
    \mathrm{.}
  \label{equ:error2}
\end{equation}

To determine $f_\mathrm{corr}$, we proceed as described in \citet{Clement} by measuring the variance of the flux measured in object-free apertures randomly positioned in an image:

\begin{enumerate}
\item For a given CFHQSIR tile, we select 2,000 positions corresponding to source-free background regions. This is achieved by selecting exclusion zones 20\arcsec\ away from the edges and all the sources detected in the image.
\item With SExtractor, we measure the flux at these positions for aperture diameters ranging from two to 14 pixels (i.e. from 0\farcs37 to 2\farcs6).
\item We compute the rms of the flux in these apertures, $ \sigma_{\Phi}$, and compare with the average of the errors reported by SExtractor, $<\sigma_{\Phi_\mathrm{SE}}>$.
\end{enumerate}

\begin{figure}\centering
    \includegraphics[width=0.5\textwidth]{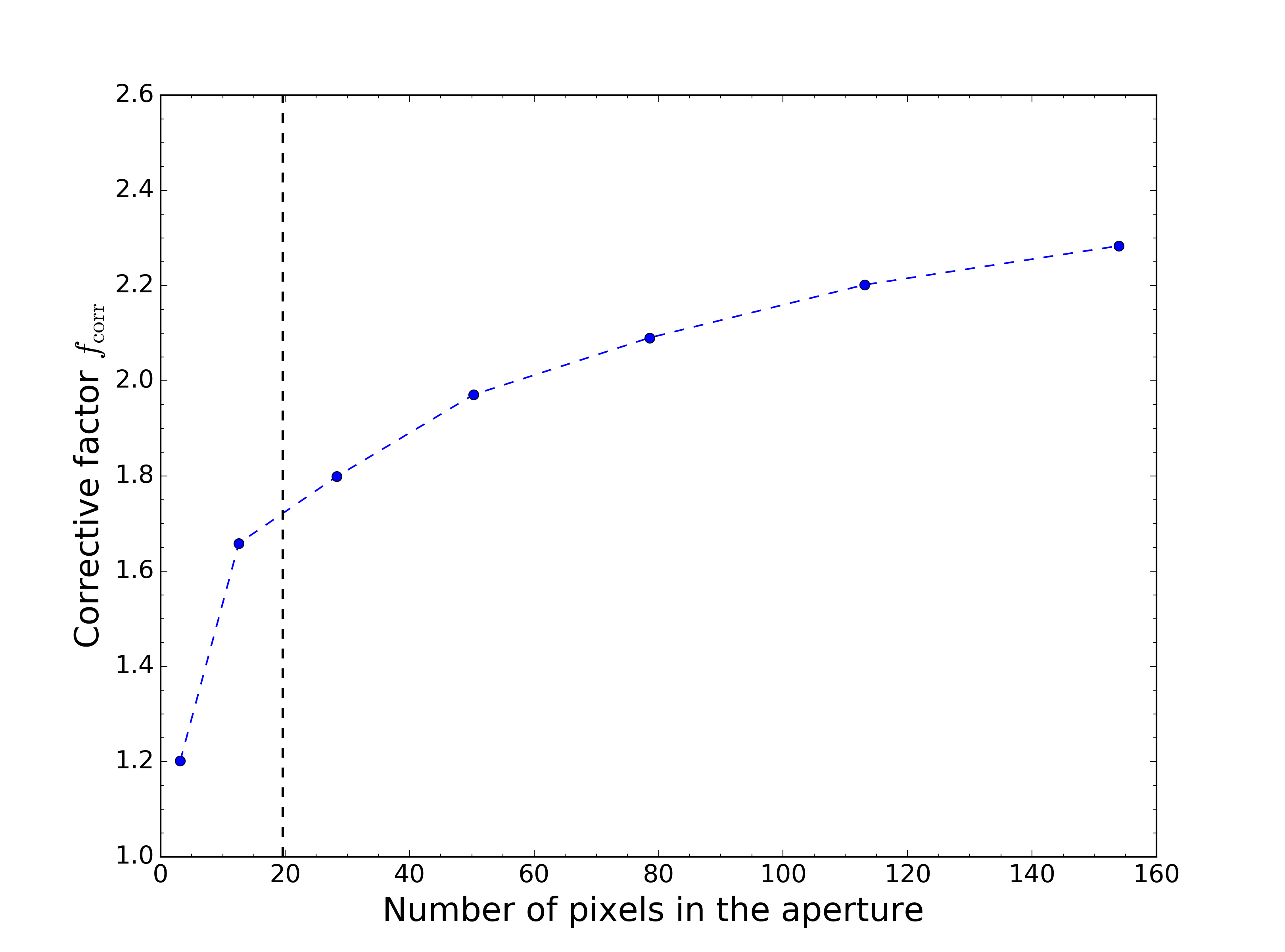}
    \caption{Corrective noise factor as a function of number of pixels in the aperture. There is a strong increase at small scales, comparable with the size of the interpolation kernel. The factor at large scales is approximately proportional to $N_\mathrm{pix}^{0.1}$. The vertical dashed line indicates the aperture diameter of $0\farcs93$ adopted in Section \ref{sec:aper}.}
      \label{fig:fig_fcorr}
\end{figure}

As expected, the error measured by SExtractor, $<\sigma_{\Phi_{SE}}>$, is proportional to $N_\mathrm{pix}^{0.5}$, while $ \sigma_{\Phi}$ evolves as $N_\mathrm{pix}^{0.6}$ at large $N_{pix}$ values. Figure~\ref{fig:fig_fcorr} shows the evolution of the ratio of these two quantities $f_\mathrm{corr} =  \frac{\sigma_{\Phi}}{<\sigma_{\Phi_\mathrm{SE}}>}$ with the number of pixels in the aperture. There are two regimes in this figure, one of steep rise at low $N_\mathrm{pix}$ values, as expected from the strong correlation of the noise between adjacent pixels due to the length of the resampling interpolation kernel, and a smoother evolution, proportional to $\sim N_\mathrm{pix}^{0.1}$ for larger $N_\mathrm{pix}$ values. As an illustration, a corrective factor of about two must be applied for aperture diameters of $1\farcs60$ corresponding to $\approx 60$ pixels. For the aperture diameter adopted in Section \ref{sec:aper} ($0\farcs93$, corresponding to $\approx 20$ pixels), the corrective factor is around $1.7$, as indicated by the vertical dashed line in Fig.~\ref{fig:fig_fcorr}.

\subsubsection{Aperture photometry}\label{sec:aper}

We first choose an aperture size around point sources to measure the sensitivity of our images. The best aperture size maximizing the signal-to-noise ratio is discussed for instance in \citet{mighell}, who recommends an aperture diameter of $\approx 2 \times$ FWHM. The optimum diameter for a circular aperture, uncorrelated noise and a gaussian PSF is $\approx 1.36 \times$ FWHM. Systematic photometric errors increase at both small and large aperture sizes ; at small aperture sizes, the aperture correction is sensitive to IQ variations within an IQ bin and errors in estimating the local background translate into large photometric errors at large aperture sizes.

We build stacks of unsaturated stars after normalization to their total flux. We then measure the curves of growth of these images for apertures ranging from one to 100 pixels in diameter in $\approx 0\farcs1$ image quality (IQ) bins. We perform a variety of tests to check that the resulting stacks and curves of growth are stable, within the same IQ bin, from images to images, or from field to field. From these curves we derive the aperture corrections for each IQ bin and measure the aperture sizes maximizing the signal-to-noise ratio. For each IQ bin we find optimum aperture diameters (assuming noise-free aperture corrections) between 1.2 and 1.4 times the PSF FWHM.
We finally adopt an aperture diameter of $0\farcs93$, which maximizes the signal-to-noise ratio in the $0\farcs7$ IQ bin, which is approximately the mean IQ value of the data.

\subsubsection{Completeness and limiting magnitudes}
\begin{figure}\centering
    \includegraphics[width=0.5\textwidth]{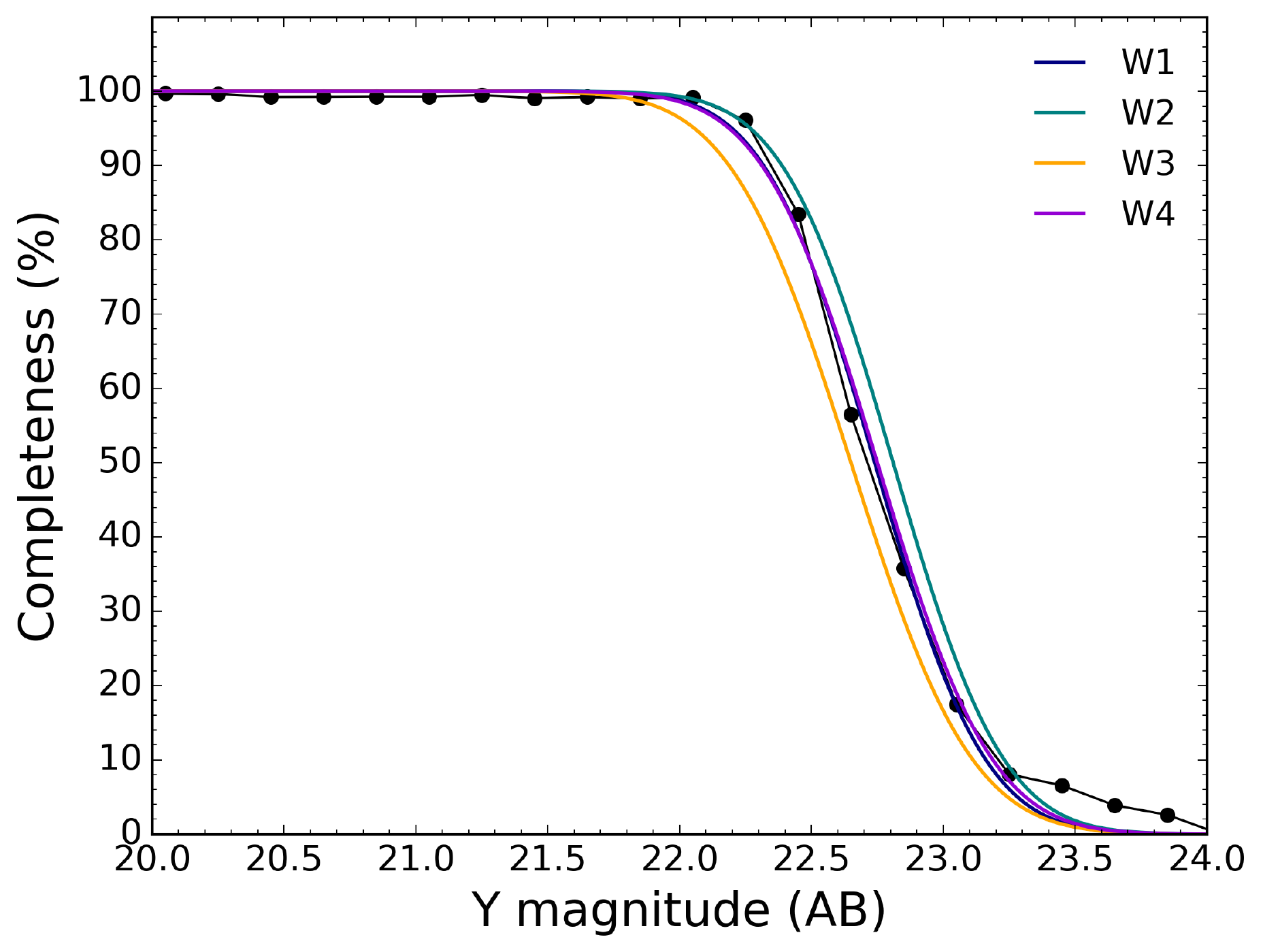}
  \caption{Point source completeness rate as a function of magnitude for each of the four CFHQSIR fields. The coloured lines correspond to fits of the data points with complementary error functions and the black filled circles are the data points for the W1 field.}
  \label{fig:completeness}
\end{figure}

We estimate the completeness of our data by randomly generating artificial sources of known magnitudes with the local PSF associated to each one-ninth of a tile corresponding to a WIRCam stack. The sources are randomly added in any region of the image that sees the sky: the gaps between the detectors are excluded, regions where there are objects, including bright stars, are not. In Fig.~\ref{fig:completeness} we plot the fraction of sources that are detected as a function of the magnitude. The magnitude limit at 80\% completeness is of the order of 22.5.

We also derive limiting magnitude maps by calculating the 5$\sigma$ limiting magnitude from the noise maps generated by SExtractor. For a given aperture, the magnitude for a detection limit at 5$\sigma$ is defined by:

\begin{equation}
     m_{5\sigma} = -2.5 \log_{10}(5 \times f_\mathrm{corr} \times \sigma_\mathrm{pix} \times  \sqrt{N_\mathrm{pix}}) - \delta_{m_\mathrm{ap}} + ZP,
    \end{equation}
 where:
 \begin{itemize}
   \item $f_\mathrm{corr}$ is the corrective noise factor determined in section~\ref{sec:error},
   \item $ \sigma_\mathrm{pix}$ is the local pixel to pixel standard deviation of the background signal as measured by SExtractor,
   \item $N_\mathrm{pix}$ is the number of pixels in the aperture,
   \item $\delta_{m_\mathrm{ap}}$ is the aperture correction.
 \end{itemize}

\begin{figure}\centering
    \includegraphics[width=0.5\textwidth]{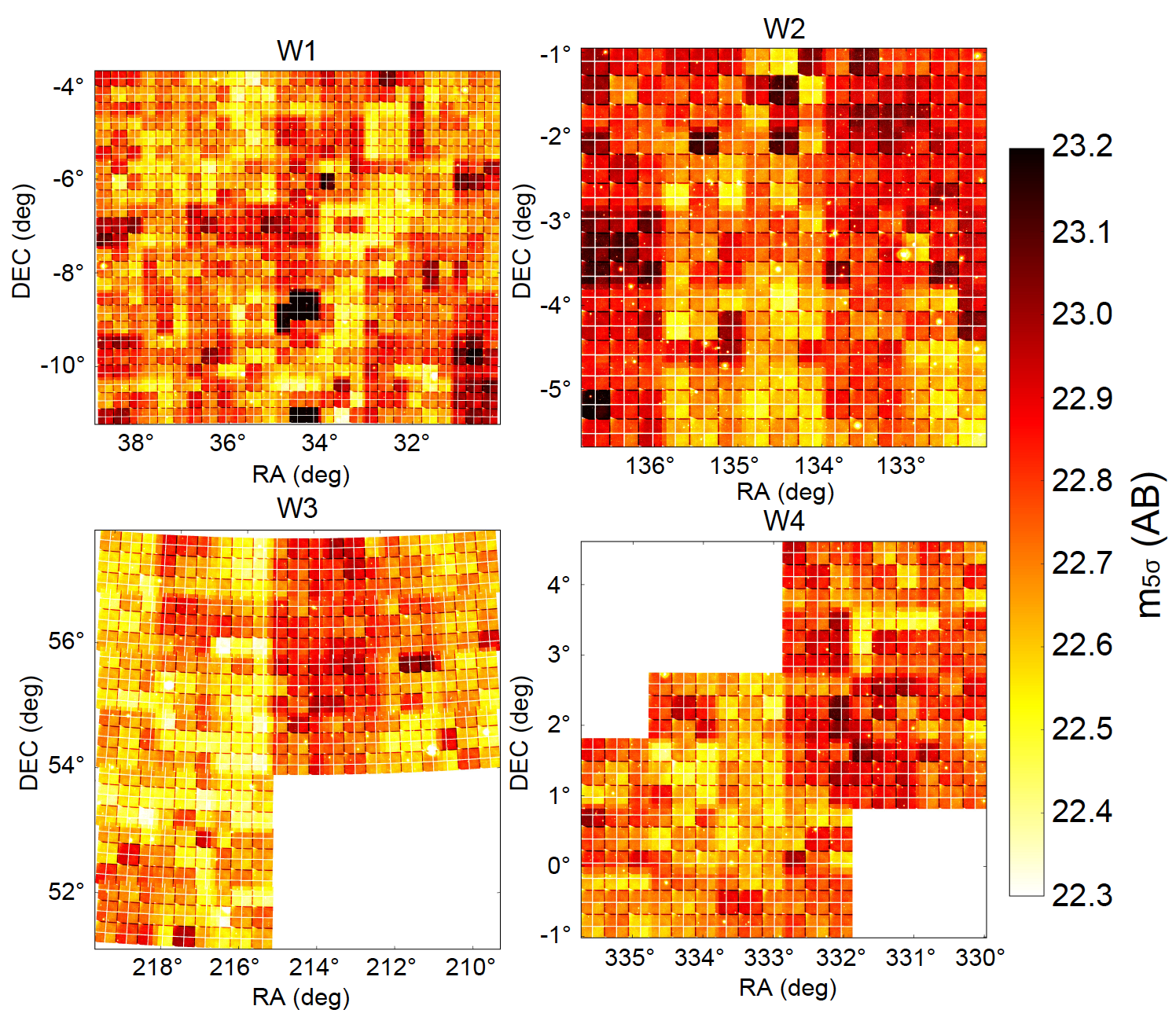}
  \caption{Sensitivity maps at 5$\sigma$ for point-like sources and $0\farcs93$ diameter apertures. The average 5$\sigma$ limiting magnitude (AB) is $\sim 22.7$, as indicated in Table~\ref{tab:cumulative_areas}.}
  \label{fig:sensitivity_maps}
\end{figure}

\begin{figure}\centering
    \includegraphics[width=0.5\textwidth]{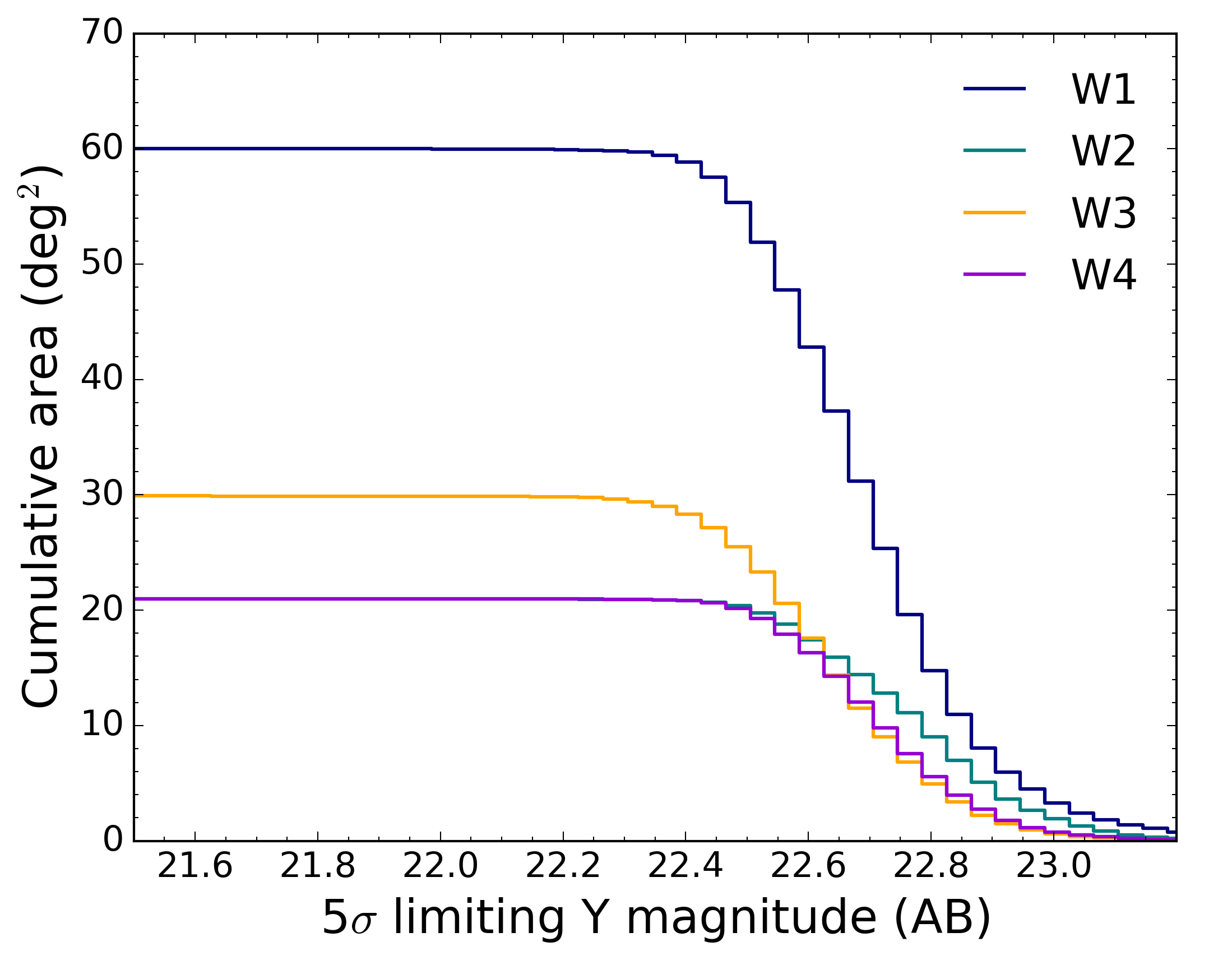}
  \caption{Cumulative area covered by the four CFHQSIR fields as a function of the 5$\sigma$ limiting magnitude.}
  \label{fig:cumulative_areas}
\end{figure}

The resulting maps are presented in Fig.~\ref{fig:sensitivity_maps}. We subsequently derived the effective areas in each of the three fields as a function of flux (or magnitude) by counting the number of pixels below a given flux. This is shown in Fig.~\ref{fig:cumulative_areas} as a function of the magnitude. We report in Table~\ref{tab:cumulative_areas} the effective areas and limiting magnitudes, with their dispersions, for the four CFHQSIR fields.

\begin{table}
\caption{CFHQSIR limiting magnitudes and areas: footprint areas of the four CFHTLS fields, effective areas with an AB limiting magnitude of $\sim$ 22.4, average limiting magnitude and dispersion.}           
\label{tab:cumulative_areas}
\centering
\begin{tabular}{c c c c c}
\hline
Field & Footprint  & Effective & 5$\sigma$ Limiting  & Error \\   
 &  area (deg$^2$) &  area (deg$^2$) &  magnitude  &  \\   
\hline
W1 & 63.8 & 60 & 22.7 & 0.15 \\
W2 & 22.5 & 21 & 22.8 & 0.16\\
W3 & 44.2$^1$ & 30 & 22.7 & 0.15 \\
W4 & 23.3 & 21 & 22.7 & 0.14  \\
\hline
\end{tabular}
\\
$^1$ This is the CFHTLS W3 footprint area. The footprint area covered by CFHQSIR in the Y-band is 32.4 deg$^2$. Here the error refers to the uncertainty in the $5\sigma$ limiting magnitude.
\end{table}

\section{The CFHQSIR catalogue}
\label{sec:catalogue}
      
In this section we present preliminary analyses performed on a CFHTLS-matched catalogue of CFHQSIR objects. To generate this catalogue, we use the seventh and final release (T0007) of the CFHTLS produced by Terapix\footnote{\url{http://terapix.iap.fr/cplt/T0007/doc/T0007-doc.html}}. These catalogues are produced in each u*, g', r', i', and z' band from a g-r-i combined "chi2" detection image for each one-square-degree CFHTLS tile. We similarly produce the Y-band matched catalogue by using these $gri-chi2$ detection images and the same SExtractor configuration parameters used for the CFHTLS catalogues. We then match this catalogue of objects covered by CFHQSIR with the CFHTLS catalogues and produce a final photometric catalogue in the u*-, g'-, r'-, i'-, z'-, and Y-bands.\\
We use this catalogue to perform a number of analyses aimed at assessing the scientific performance of the CFHQSIR data. We compare the CFHQSIR data with the (deeper) Vista Deep Extragalactic Observations survey (VIDEO) survey data for number counts and completeness (Sect. \ref{sec:Ncounts}), and we evaluate the benefit of the Y-band data in the determination of photometric redshifts (Sect. \ref{sec:zphot})

\subsection{Comparison with the VIDEO survey}\label{sec:Ncounts}

\begin{figure}\centering
        \includegraphics[width=0.4\textwidth]{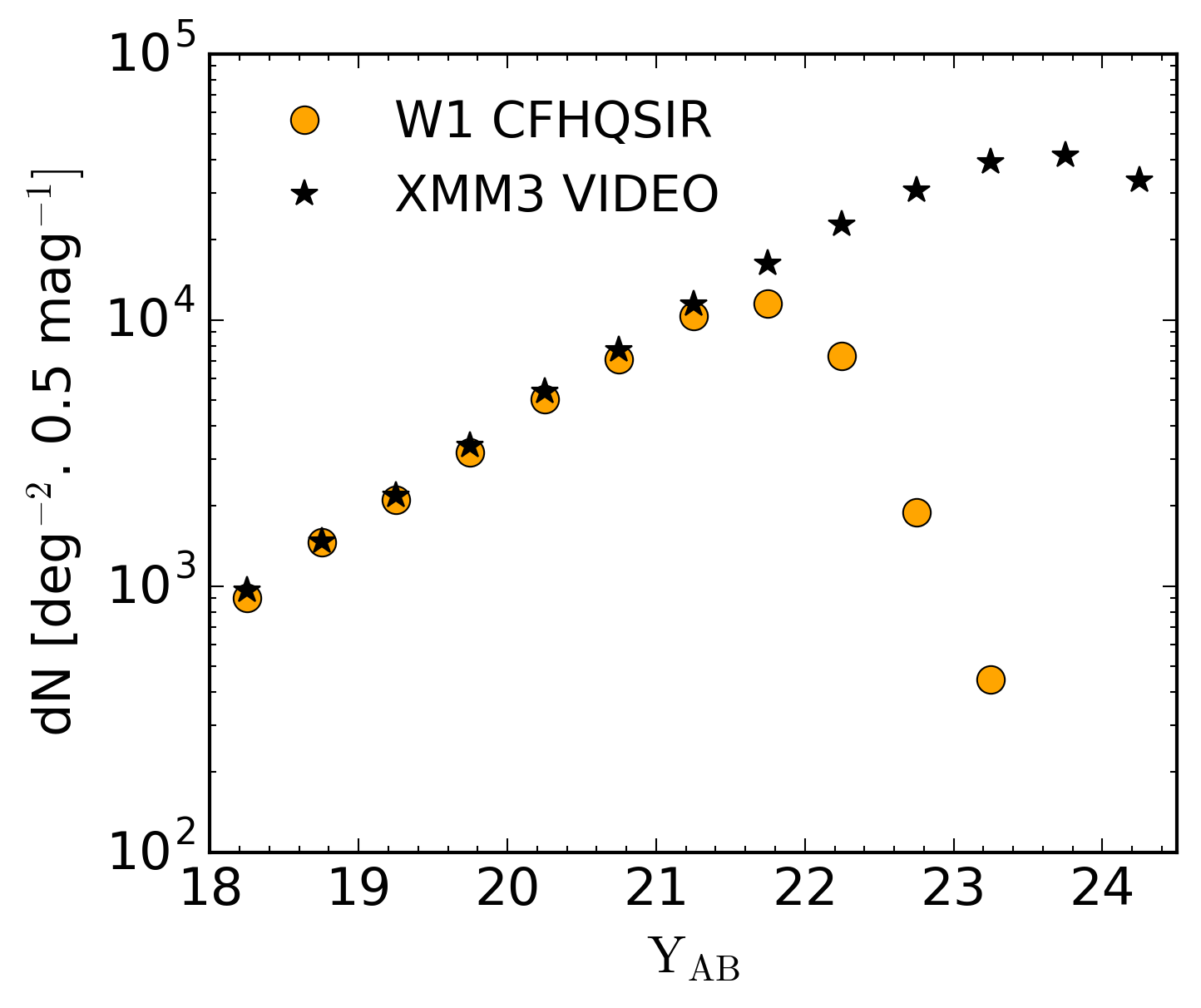}
        \caption{CFHQSIR number counts (orange points) and VIDEO number counts \citep[black stars, ][]{Jarvis2013} in the XMM3 sub-field of the W1 field. The Poisson errors are smaller than the symbol sizes. The magnitudes are the \texttt{MAG\_AUTO} CFHQSIR and VIDEO magnitudes.}
        \label{fig:Nbcounts}
\end{figure}

For the sake of assessing the performance of the CFHQSIR survey, we compare the CFHQSIR and VIDEO \citep[VISTA Deep Extragalactic Observations Survey; ][]{Jarvis2013} number counts. VIDEO is approximately two magnitudes deeper than CFHQSIR albeit over a much reduced area. We use the XMM1 and XMM3 fields, which are both in the W1 footprint and cover a total area of three square degrees. The VIDEO data in these fields have a $5\sigma$ depth of Y $\sim 24.9$ in 2\arcsec\ diameter apertures. We use the SExtractor \texttt{MAG\_AUTO} magnitudes reported in the fourth data release (DR4) of the VIDEO catalogue. Because our catalogue is optically selected, we apply a cut in SNR in the Y-band in order to keep objects well detected in this band. After visual and qualitative tests we adopt SNR $\geq$ 6. Figure~\ref{fig:Nbcounts} shows the CFHQSIR and VIDEO Y-band number counts in the same sub-field (XMM3) of the W1 field. \\

\begin{figure}\centering
	\includegraphics[width=0.5\textwidth]{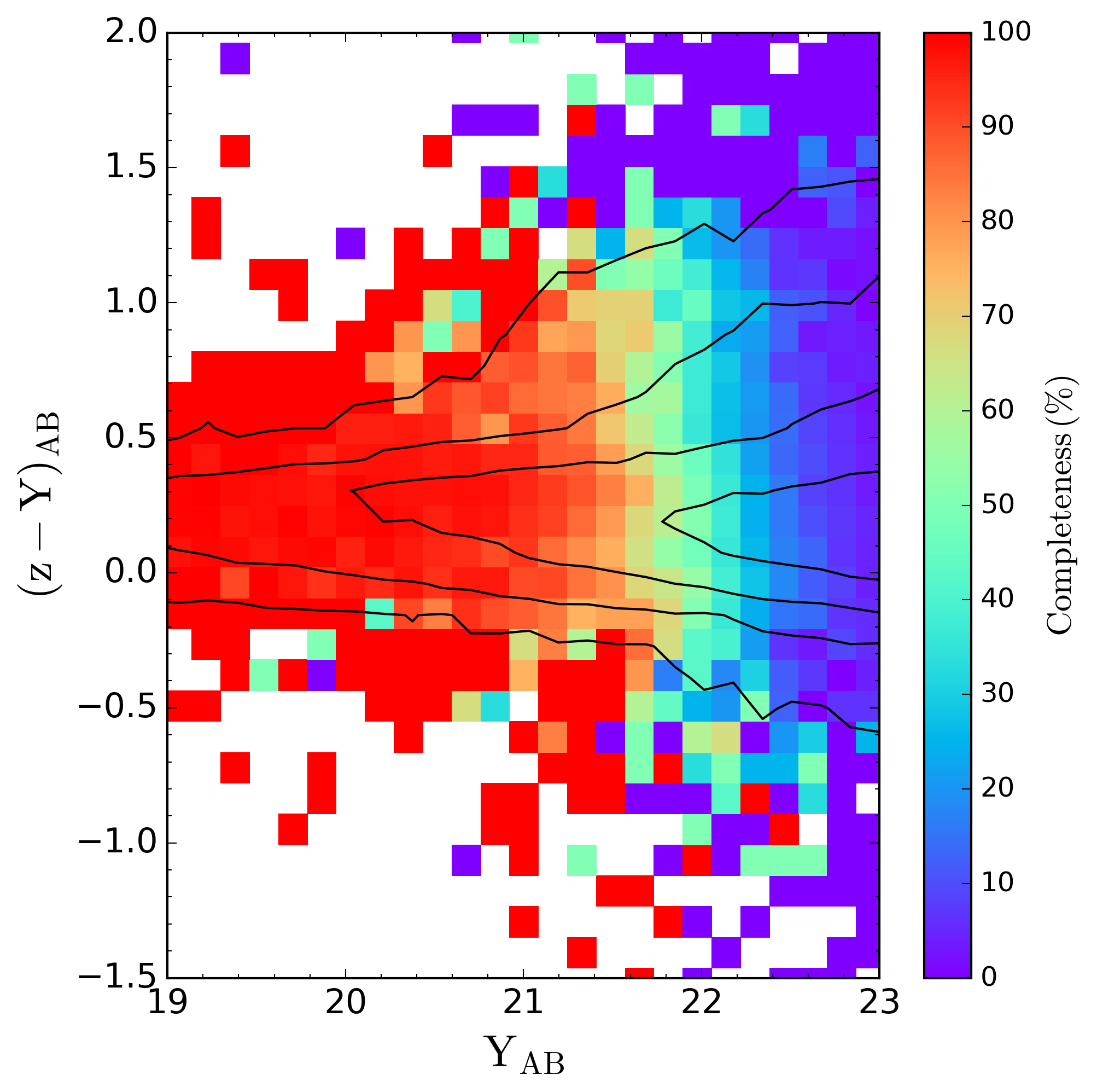}
	\caption{Bidimensional completeness of the CFHQSIR survey relative to the VIDEO survey as a function of the VIDEO Y magnitude and (z-Y) colour. From the outer one to the inner one, the black solid lines represent the iso-contours corresponding to 99\%, 96\%, 86\%, 69\% and 20\% of all VIDEO sources.}
	\label{fig:completudeVIDEO}
\end{figure}

Considering that our catalogue is optically-selected and that objects only seen in the Y-band may be missing, we investigate the completeness and the comparison with the VIDEO survey in more detail. To this aim, we match both catalogues with a 1\farcs0 search radius. We then compute the fraction of VIDEO sources present in our catalogue as a function of the Y-band (VIDEO) magnitude and (z-Y) colour. The results are shown in Fig.~\ref{fig:completudeVIDEO}, where the colour code reflects the completeness rate. As expected, the incompleteness increases with magnitude at Y $\gtrsim 22.0$ and increases from the bluer to the redder objects.

\subsection{Photometric redshifts} \label{sec:zphot}
 
Following \cite{Erben2013} and \cite{Moutard2016}, we build a multi-wavelength dataset based on SExtractor isophotal apertures (corresponding to SExtractor \texttt{MAG\_ISO}). By construction, \texttt{ISO} apertures contain pixels characterized by a high signal-to-noise ratio, which tends to make them better matched to the shape of the sources than traditional circular apertures while delivering less noisy measurements than Kron-like apertures (corresponding to SExtractor \texttt{MAG\_AUTO}). This leads to more precise measurements of the colours and, therefore, more accurate photometric redshifts \citep{Hildebrandt2012}.

Following the method developed by \cite{Moutard2016}, we compute the total magnitude for each source by rescaling the \texttt{MAG\_ISO} magnitudes to the \texttt{MAG\_AUTO} magnitude. We apply the following formula for each band $b =$ (u*, g', r', i', z', Y): $m_{\mathrm{total},b} = m_{\mathrm{ISO},b} + \delta_m$, where $\delta_m$ is the scaling factor, defined as

\begin{equation}
\delta_{m} = \frac{\sum_{f} (m_{\mathrm{AUTO},f}-m_{\mathrm{ISO},f}) \times w_{f}}{\sum_{f} w_{f}}
\mathrm{,}
\end{equation}
where $f$ refers to the g', r', i' and Y filters and the weight $w_{f}$ is given by $1/w_{f} = (\sigma^{2}_{\mathrm{AUTO},f} + \sigma^{2}_{\mathrm{ISO},f})$.

All magnitudes are then corrected from the Galactic extinction using the tabulated values of the reddening excess $E(B-V)$ in the CFHTLS catalogues for each source and the Cardelli law \citep{Cardelli1989}. To account for the effects of correlated noise described in Sect. \ref{sec:error}, we apply an empirically-derived corrective noise factor of 1.9 to the \texttt{ISO} magnitude errors. This factor does provide on average the best confidence levels when estimating the photometric redshifts.

The photometric redshifts were computed with the SED-fitting (Spectral Energy Distribution) code \lephare \citep{Arnouts1999, Ilbert2006}, by following the very same scheme as \citet{Moutard2016} (please refer to this paper for further details). In brief, we made use of the SED template library from  \citet{Coupon2015}, by considering extinction laws from \citet{Prevot1984} for evolved spiral templates and based on \citet{Calzetti2000} for star-burst templates, while no extinction was considered for elliptical templates. 
Any systematic disagreement between the photometry and the template library was corrected with \lephare via the method described by \citet{Ilbert2006}, that is, by tracking a systematic shift between predicted and observed magnitudes in each band at a given redshift.

For this purpose, as well as for testing the association of CFHQSIR and CFHTLS observations for the derivation of photometric redshifts, we collated spectroscopic redshifts from the literature to build a sample of more than 76,000 of the most secure redshift measurements. Namely, we made use of
i) the $i < 19.9$-limited Sloan Digital Sky Survey \citep[SDSS,][]{York2000} and Baryon Oscillation Spectroscopic Survey \citep[BOSS,][]{Dawson2013},
ii) the VIMOS (Visible Multi-Object Spectrograph) Very Deep Survey \citep[VVDS,][]{LeFevre2013} "Wide" ($i$ < 22.5 in W4) and "Deep" ($i$ < 24 in W1),
iii) the second public data release (PDR2) of the VIMOS Public Extragalactic Redshift Survey \citep[VIPERS; ][]{Scodeggio2017}, which provides a $i$ < $22.5$-limited sample of spectroscopic redshifts covering $0.5 < z < 1.2$ with a sampling rate of $40\%$ over 24 deg$^2$ in W1 and W4, 
iv) the redshifts from the $K < 23$-limited UKIDSS spectroscopic Ultra Deep Survey \citep[UDSz,][]{Bradshaw2013,McLure2013} when available in W1.

\begin{figure*}[t]
	\includegraphics[width=\textwidth]{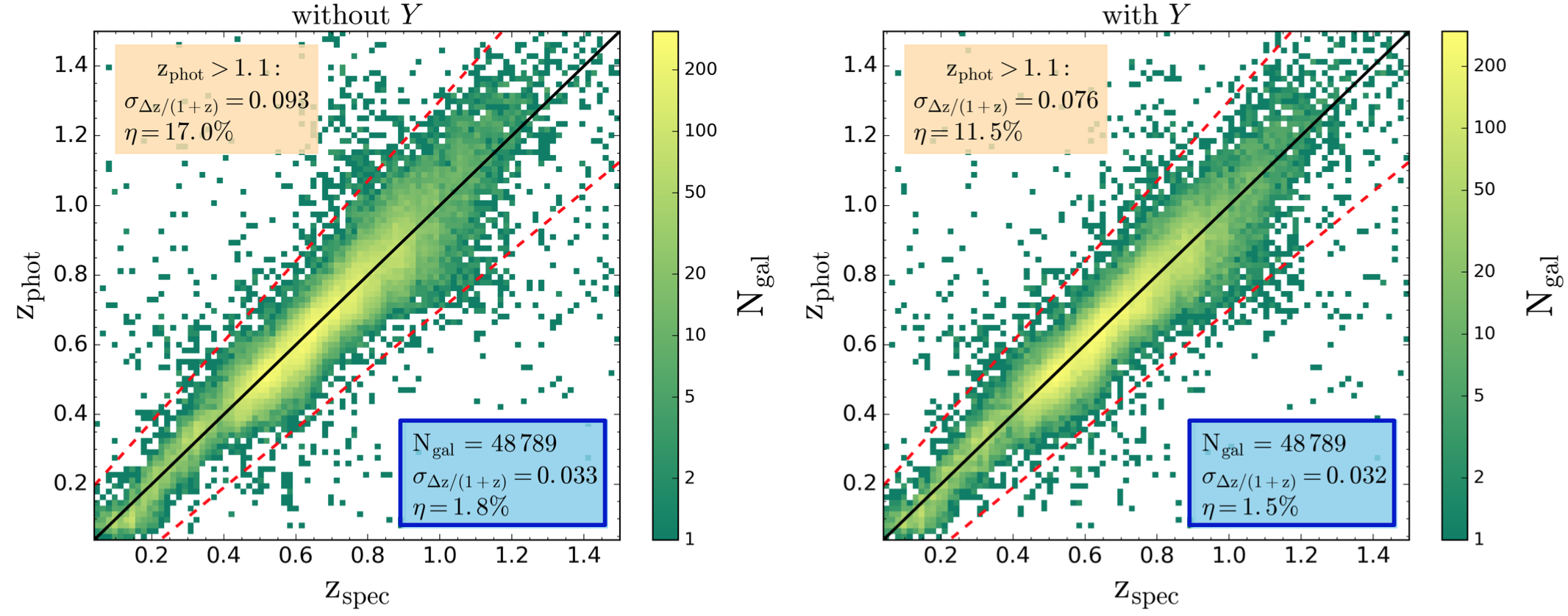}
	\caption{Comparison between the photometric and spectroscopic redshifts of our sample, with i'-band magnitude at $15.0 <$ i' $< 25.0$. The red dashed lines correspond to the outlier limit beyond which $\sigma_{\Delta z/(1+z)} > 0.15$.}
		\label{fig:zp_zs}
\end{figure*}

Figure \ref{fig:zp_zs} shows the comparisons between 
spectroscopic and photometric redshifts, as derived with CFHTLS observations only (left panel) or when combining CFHTLS and CFHQSIR Y-band observations (right panel). 

The addition of the Y-band marginally improves the normalized median absolute deviation (NMAD) used to define the scatter\footnote{$\sigma_{\Delta z/(1+z)} = 1.4821 \times median(~|z_{phot}-z_{spec}| / (1+z_{spec})~).$}, from $\sigma_{\Delta z/(1+z)} =  0.033$ to $\sigma_{\Delta z/(1+z)} =  0.032$. The improvement is more significant on the outlier rate\footnote{$\eta$ being the fraction of galaxies with $|z_{phot}-z_{spec}|/(1+z_{spec})>0.15.$}, which decreases from $\eta = 1.8\%$ to $\eta = 1.5\%$ when adding the Y-band.

Focusing on photometric redshifts $z_{phot} > 1.1$, the contribution of the Y-band is even more obvious, with a drop of the outlier rate from $\eta = 17.0\%$ to $\eta = 11.5\%$ when combining CFHTLS and CFHQSIR data. Significant as well is the increase of the photometric redshift precision at $z_{phot} > 1.1$, with a dispersion reduced from $\sigma_{\Delta z/(1+z)} = 0.093$ to $\sigma_{\Delta z/(1+z)} = 0.076$. 

 \begin{figure}\centering

	\includegraphics[width=\columnwidth]{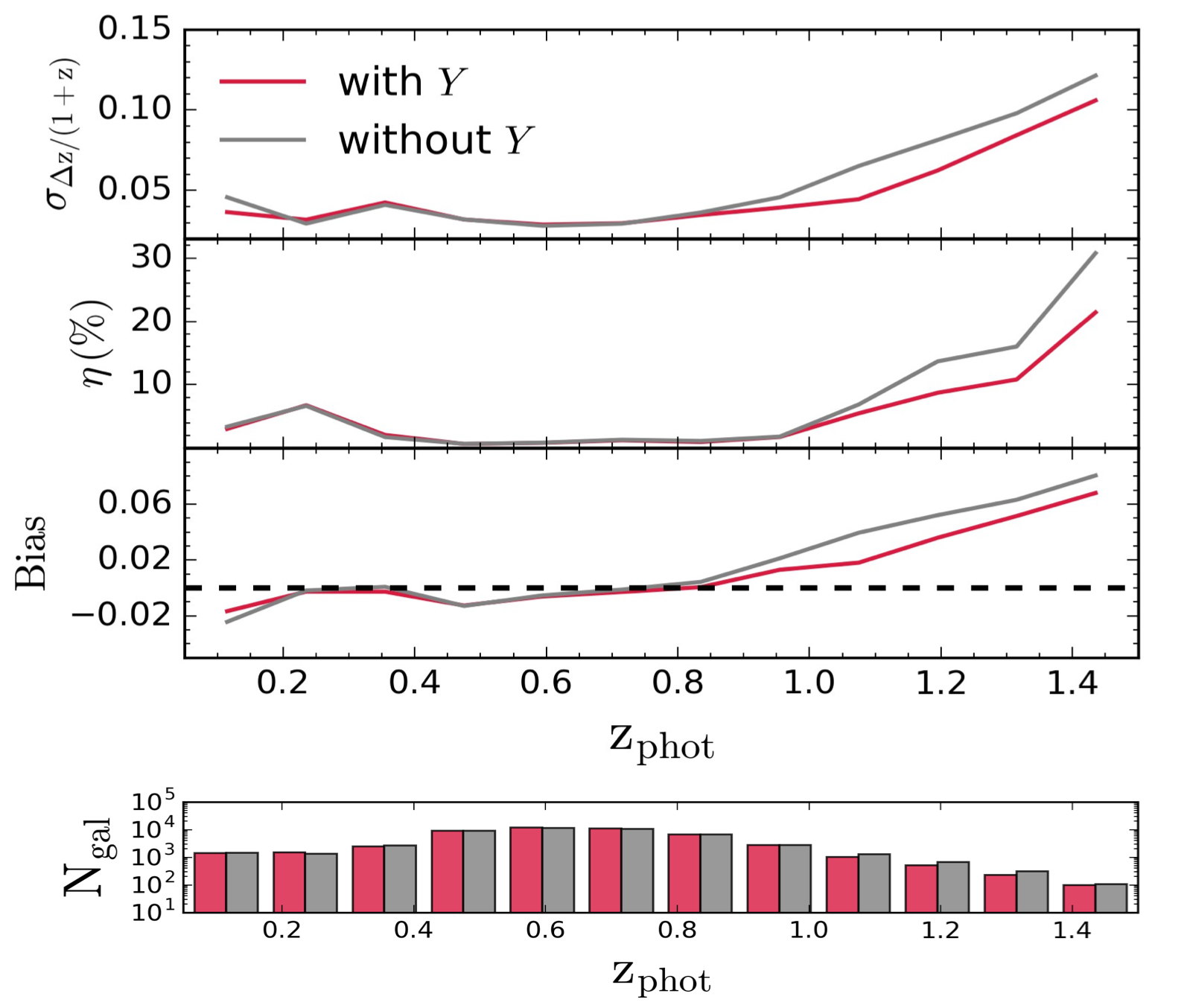}
	\caption{Photometric redshift statistics of our sample, as a function of photometric redshift. The red and grey colours refer respectively to the cases with and without considering the CFHQSIR Y-band. The sample is limited to the $i'$-band magnitude at $15.0 < i' < 25.0$.}
	\label{fig:zphot_stats}
\end{figure}

Figure \ref{fig:zphot_stats} displays the dispersion $\sigma_{\Delta z/(1+z)}$, the outlier rate $\eta$, and the redshift bias\footnote{Where the bias, defined by $(z_{phot}-z_{spec}) / (1+z_{spec})$ indicates the inclination of photometric redshifts to be systematically under- or over-estimated} as a function of the photometric redshift, as well as the number of galaxies $\mathrm{N_{gal}}$ used in each redshift bin. The addition of the Y-band clearly improves the overall robustness of the photometric redshifts at $z \gtrsim 1.0$ despite the shallow CFHQSIR sensitivity. For instance, the redshift range where $\sigma_{\Delta z/(1+z)} < 0.06$ is extended from $z \simeq 1.05$ to $z \simeq 1.2$, where the use of the Y-band maintains an outlier rate $\eta < 8.8\%$ against $\eta < 13.8\%$ without the Y-band. The CFHQSIR data therefore extends the effective redshift range probed by the CFHTLS for studies based on photometric redshifts, for example galaxy-evolution or weak-lensing analyses.

\section{Conclusion}
We have conducted a new Y-band imaging survey covering the four fields of the CFHTLS-Wide, as part of the CFHQSIR. The average image quality of the data is 0\farcs70, with a standard deviation of 0\farcs08. The observations reach a limiting magnitude Y$_\mathrm{AB} \sim 22.4$ ($5\sigma / 0\farcs93$)
over 130 square degrees. \\
We combined the CFHTLS-T0007 optical data and the CFHQSIR wide band data to produce a multi-wavelength catalogue. We found good agreement between the number counts from this catalogue and the deeper VIDEO data up to a magnitude of Y$_\mathrm{AB} \sim 21.3$ at SNR $\geq 6$. \\
We tested the improvement of the CFHTLS photometric redshifts with the addition of the Y-band. We derived photometric redshifts for more than 40,000 sources in the W1 and W4 fields and compared them to their spectroscopic redshifts. We found that the addition of the Y-band improves the accuracy of photometric redshifts at $z \gtrsim 0.9$.\\
The images as both WIRCam stacks and one-square-degree tiles, as well as the CFHQSIR catalogue including $\sim 8.6$ million sources, are accessible at the following link: \url{http://apps.canfar.net/storage/list/cjw/cfhqsir} (see Appendix \ref{appendice} for more details).


\begin{acknowledgements}
	We thank the CFHT staff, and in particular Todd Burdullis, QSO Operations Specialist, and Karun Thanjavur, previously resident astronomer at CFHT, for their dedication and efforts in accompanying the CFHQSIR observations and early data reduction. This paper uses data from the VIMOS Public Extragalactic Redshift Survey (VIPERS). VIPERS has been performed using the ESO Very Large Telescope, under the “Large Programme” 182.A-0886. The participating institutions and funding agencies are listed at \texttt{\url{http://vipers.inaf.it}}. This paper uses data from the VIMOS VLT Deep Survey (VVDS) obtained at the ESO Very Large Telescope under programmes 070.A-9007 and 177.A-0837, and made available at the CeSAM data centre, Laboratoire d’Astrophysique de Marseille, France. Funding for SDSS-III has been provided by the Alfred P. Sloan Foundation, the Participating Institutions, the National Science Foundation, and the US Department of Energy Office of Science. The Participating Institutions of the SDSS-III Collaboration are listed at \texttt{\url{http://www.sdss3.org/}}. We thank the anonymous referee for constructive comments that have helped us improve the quality of the paper.
	
\end{acknowledgements}


\bibliographystyle{aa} 
\bibliography{biblio} 

\begin{appendix}
	\section{The CFHQSIR data release}
	\label{appendice}
	
	The CFHQSIR data release is public and accessible at the following link: \url{http://apps.canfar.net/storage/list/cjw/cfhqsir}. The data are:
	\begin{itemize}
		\item 1445 WIRCam stacks of 0.1 deg$^{2}$ and their corresponding weight-maps (\texttt{FITS} images):
		\medskip
		\begin{itemize}
			\item 648 for W1,
			\item 225 for W2,
			\item 347 for W3 and
			\item 225 for W4;
		\end{itemize}
		\bigskip
		\item 163 tiles of 1 deg$^{2}$ and their corresponding weight-maps (\texttt{FITS} images):
		\medskip
		\begin{itemize}
			\item 72 for W1,
			\item 25 for W2,
			\item 41 for W3 and
			\item 25 for W4;
		\end{itemize}
		\bigskip
		\item a u*-, g'-, r'-, i'-, z', Y-bands catalogue (\texttt{FITS} table) created by running SExtractor in dual-image mode using the CFHTLS \textit{gri - chi2} images as the reference source detection images and the Y-band CFHQSIR one-square-degree tiles as photometry images.
		
	\end{itemize}
	
	Considering the ratio of integration times in the CFHTLS and CFHQSIR images (hours versus minutes), the depth of the Y-band CFHQSIR images is much shallower than the depth of the CFHTLS images and we therefore applied a cut in magnitude to retain meaningful Y-band values. We also applied a cut in the z'-band, considering the relative depths in this band compared to the ultra-deep $gri-chi2$ detection images. We choose $[(\mathrm{z'} \leq 23.5) \lor (\mathrm{Y} \leq 23.0)]$, which matches the sensitivity limits in these two bands. Over the $\sim 130$ deg$^{2}$ covered by CFHQSIR, the resulting catalogue contains a total of $\sim 8.6$ million sources. \\
	
\end{appendix}

\end{document}